\documentclass[aps,amsmath,amssymb,prb,10pt,twocolumn,eqrefnum]{revtex4-1}
\usepackage{graphicx,txfonts,feynmp}
\usepackage{bm}
\usepackage{mathrsfs,eufrak}
\usepackage{booktabs}
\usepackage{comment}
\usepackage{mathtools}
\usepackage{appendix}
\mathtoolsset{showonlyrefs}

\newcommand{\al}{\alpha}

\newcommand{\g}{\gamma}
\newcommand{\de}{\delta}

\newcommand{\z}{\zeta}

\newcommand{\thi}{\theta}
\newcommand{\ii}{\iota}
\newcommand{\ka}{\kappa}
\newcommand{\la}{\lambda}

\newcommand{\p}{\pi}

\newcommand{\s}{\sigma}

\newcommand{\f}{\phi}

\newcommand{\w}{\omega}
\newcommand{\W}{\Omega}

\newcommand{\G}{\Gamma}
\renewcommand{\S}{\Sigma}

\newcommand{\y}{\psi}

\newcommand{\pd}{\partial}

\newcommand{\round}[1]{\left( #1 \right)}
\renewcommand{\square}[1]{\left[ #1 \right]}

\newcommand{\mat}[4]{\left(\begin{array}{cc}#1&#2\\#3&#4\end{array}\right)}

\newcommand{\beq}{\begin{equation}}
\newcommand{\eeq}{\end{equation}}
\newcommand{\Beq}{\begin{eqnarray}}
\newcommand{\Eeq}{\end{eqnarray}}
\newcommand{\bml}{\begin{multline}}

\newcommand{\bea}{\begin{align}}
\newcommand{\ena}{\end{align}}
\newcommand{\bsp}{\begin{split}}
\newcommand{\esp}{\end{split}}
\newcommand{\down}{\downarrow}
\newcommand{\up}{\uparrow}

\newcommand{\csch}{\mbox{csch}}

\newcommand{\bS}{{\boldsymbol{S}}}

\newcommand{\bi}{{\boldsymbol i}}

\newcommand{\bj}{{\boldsymbol j}}

\newcommand{\bk}{{\boldsymbol k}}

\newcommand{\bq}{{\boldsymbol q}}

\newcommand{\hG}{\hat G}
\newcommand{\hg}{\hat g}

\DeclareMathOperator{\tr}{Tr}

\newcommand{\sH}{h}
\newcommand{\sA}{\mathscr{A}}
\newcommand{\cA}{\mathcal{A}}
\newcommand{\cB}{\mathcal{B}}

\newcommand{\cS}{\mathcal{S}}

\newcommand{\hD}{\hat{D}}
\newcommand{\sD}{\mathcal{D}}

\newcommand{\sV}{\mathscr{V}}
\newcommand{\sN}{\mathcal{N}}

\newcommand{\bx}{\boldsymbol{x}}

\newcommand{\ve}{\varepsilon}

\newcommand{\bzero}{{\boldsymbol 0}}

\newcommand{\bR}{{\boldsymbol R}}

\newcommand{\tSig}{\tilde{\Sigma}}

\begin{document}
\title{Spin transport in an electrically-driven magnon gas near Bose-Einstein condensation:\\
Hartree-Fock-Keldysh theory}
\date{\today}

\author{So Takei}
\email{Email: stakei@qc.cuny.edu}
\affiliation{Department of Physics, Queens College of the City University of New York, Queens, NY 11367, USA and \\ 
Physics Doctoral Program, Graduate Center of the City University of New York, New York, NY 10016, USA}

\begin{abstract}
An easy-plane ferromagnetic insulator in a uniform external magnetic field and in contact with a phonon bath and a normal metal bath is studied theoretically in the presence of dc spin current injection via the spin Hall effect in the metal. The Keldysh path integral formalism is used to model the magnon gas driven into a nonequilibrium steady state by mismatched bath temperatures and/or electrical injection, and we analyze the magnon system in the normal (uncondensed) state, but close to the instability to Bose-Einstein condensation (BEC), within the self-consistent Hartree-Fock approximation. We find that the steady state magnon distribution function generally has a non-thermal form that cannot be described by a single effective chemical potential and effective temperature. We also show that the BEC instability in the electrically-driven magnon system is signaled by a sign change in the imaginary part of the poles for long-wavelength magnon modes and by the divergence of the nonequilibrium magnon distribution function. In the presence of two bath temperatures, we find that the correlation length of the superfluid order parameter fluctuations exhibits nontrivial finite temperature crossover behaviors that are richer than the standard crossover behaviors obtained for the vacuum-superfluid transition in an equilibrium dilute Bose gas. We study the consequences of these thermal crossovers on the magnon spin conductivity and obtain an inverse square-root divergence in the spin conductivity in the vicinity of the electrically-induced BEC instability. A spintronics device capable of testing our spin transport predictions is discussed. 
\end{abstract}

\maketitle
\unitlength=1mm

\section{Introduction}
Magnetic insulators provide an attractive arena to study nonequilibrium Bose-Einstein condensation (BEC) of magnons in a solid state environment.\cite{duineBOOK17,sunJPD17} Unlike BECs in ultra-cold atomic systems and superconductors, BECs in solids require nonequilibrium states in which external pumping with sufficient energy is necessary to compensate for incessant quasiparticle decay.\cite{dengRMP10,szymanskaBOOK11} An early spectroscopic evidence for room temperature BEC of magnons was reported in a ferrimagnetic insulator, yttrium iron garnet (YIG), driven by parametric pumping.\cite{demokritovNAT06,demidovPRL07,demidovPRL08} According to the prevailing heuristic picture, the BEC forms during a {\em transient} stage after a sufficient number of hot incoherent magnons are injected into a narrow region of the spectrum.\cite{sergaNATC14,demidovNATC17} Once the pump is switched off, the magnons rapidly thermalize to a quasi-equilibrium state via multi-magnon scattering and condense before they ultimately decay into the lattice.

An alternative pumping mechanism is dc electrical pumping. This mechanism involves interfacing the magnetic insulator to a normal metal with strong spin-orbit coupling and utilizing the spin Hall effect for spin injection.\cite{benderPRL12,benderPRB14} Exchange coupling at the interface allows for magnon injection into (absorption out of) the insulator via annihilation (creation) of spinful particle-hole excitations in the metal, so that the metal serves both as a spin injector and a dissipative environment from the viewpoint of the insulator. Here, the threshold injection strength necessary for magnon condensation can be attained by increasing the electrical current inside the normal metal. An interesting aspect of this proposal is the possibility to realize magnon BECs in a {\em nonequilibrium steady state} | a stationary flow-equilibrium state | in which spin lost into the dissipative environment is precisely compensated by spin injection from the metal.\cite{benderPRL12,benderPRB14}

Motivated by this proposal, a recent experiment achieved magnon BEC using dc electrical spin injection in a bilayer setup consisting of YIG and a heavy-element metal platinum (Pt). The experiment reported a significant increase in the two-terminal dc spin conductivity once the magnon density surpassed the critical value needed for BEC and a compelling evidence for dissipationless spin transport mediated by the magnon condensate.\cite{wimmerCM18} This exciting development calls for a detailed study of spin transport through magnon gases in the vicinity of the BEC transition and in the simultaneous presence of dissipation and electrical pumping.

On a broader note, quasiparticle BECs in driven-dissipative steady states have been studied in different contexts including photons\cite{klaersNAT10}, excitons in coupled quantum Hall bilayers\cite{eisensteinNAT04} and exciton-polaritons in light-driven semiconductor heterostructures\cite{kasprzakNAT06}. In exciton-polariton systems, the Keldysh path integral formalism has proved useful for systematically analyzing the effects of the drive and decay on the coherent state dynamics and accounting for fluctuations beyond the mean-field limit.\cite{szymanskaPRL06} This powerful theoretical machinery is now proving useful in rigorously treating magnon BECs in driven-dissipative environments as well.\cite{troncosoPRB19} Furthermore, the BEC transition of ferromagnetic exchange magnons belongs to the same universality class as the vacuum-superfluid phase transition in dilute Bose gases.\cite{sachdevBOOK11} The rigorous theoretical formulation of the nonequilibrium magnon BEC should therefore set the stage for studying how departures from thermal equilibrium and particle/energy conservation modify the behavior of Bose gases near this well-known phase transition.

In this work, we develop a Keldysh functional integral theory for a driven-dissipative easy-plane ferromagnetic insulator coupled to a bosonic (phonon) bath and a metallic bath. Both baths introduce spin/energy loss (i.e., Gilbert damping) in the ferromagnet, and a nonequilibrium spin accumulation in the metal (generated, e.g., via the spin Hall effect) allows for incoherent spin injection (dc electrical pumping) into the magnet. In this work, we allow the two baths to have two different temperatures, i.e., $T_B$ for the phonon bath and $T_F$ for the metallic bath. Starting from a full microscopic Keldysh action consisting of the magnon system and the baths, the reduced Keldysh action for the magnons alone is first obtained by tracing out the bath degrees of freedom. Using the resulting effective action, we analyze the properties of the magnons by focusing exclusively on the normal (uncondensed) phase, considering various bath temperatures and electrical pumping strengths, and accounting for the quartic magnon-magnon interaction term by generalizing the self-consistent Hartree-Fock approach used for equilibrium dilute Bose gases to the nonequilibrium.\cite{fetterBOOK71}

Before delving into the details of the work, we first present a brief summary of the main results. Throughout this work, our focus is on the asymptotic steady state of the magnon system, a state emerging after the magnons have interacted with the baths for a long time. In this steady state, we find that the magnon distribution function, for mismatched bath temperatures $T_B\ne T_F$ and/or in the presence of electrical pumping, generally has a non-thermal form that cannot be described by a single effective chemical potential and effective temperature. It is in fact determined by a superposition of the two bath distribution functions weighted by their respective coupling strengths to the magnons. This result contrasts with similar past studies of electrically pumped magnon BECs, in which magnons were assumed to be internally thermalized to the Bose-Einstein distribution function with a well-defined effective temperature and a chemical potential.\cite{benderPRB14,fjaerbuPRB17}

One important consequence of this non-thermal magnon distribution function is that it is {\em not} possible to induce a magnon BEC by elevating the metal temperature above the phonon temperature while maintaining zero electrical pumping. In other words, the formation of a magnon BEC solely via the spin Seebeck effect is not possible. This is because the magnon distribution function is given by a superposition of both bath distribution functions, so heating up the metal always leads to partial heating of the magnons and hinders condensation. This claim is essentially consistent with the analysis of Ref.~\onlinecite{benderPRB14} in their floating-magnon-temperature regime.

The presence of the two bath temperatures has notable consequences on the critical phenomena surrounding the BEC transition as well. An important aspect of the standard BEC critical phenomena in equilibrium Bose gases is the finite temperature crossover behavior exhibited by the correlation length for the superfluid order parameter fluctuations.\cite{sachdevBOOK11} For the standard Bose gas in equilibrium, the different finite temperature regimes are defined in terms of a single temperature $T$, i.e., the magnon temperature. However, the nonequilibrium magnons in the current work are defined in terms of two bath temperatures $T_B$ and $T_F$. We therefore find crossover behaviors with respect to both bath temperatures that are richer than those obtained in the standard counterpart. In particular, since both baths can contribute to the thermal magnon population in the ferromagnet, both can act to cut off the correlation length. Which bath temperature ultimately cuts off the correlation length in the high temperature regime depends on the relative magnitudes of the temperatures and how strongly the baths hybridize with the magnons.

In the latter part of the work, we evaluate the dc spin conductivity of the magnon gas using Kubo formalism, and find that the above-mentioned finite temperature crossovers in the correlation length are reflected in the conductivity. For equal bath temperatures $T_B=T_F\equiv T$ and zero electrical pumping, for example, the dc conductivity traverses through three different temperature regimes, each possessing a distinct temperature dependence, as $T$ is increased from absolute zero. When the temperature is much lower than the bare magnon gap $\mu_0$ (defined, e.g., by the intrinsic magnetic anisotropy of the ferromagnet and the external magnetic field), thermal magnon population is exponentially suppressed and the dc conductivity obeys $\s_0\propto T^{3/2}e^{-\mu_0/k_BT}$. However, as the bath temperature increases, magnon population increases correspondingly and the conductivity crosses over to algebraic scaling $\s_0\propto T$. For yet higher temperatures, where the correlation length is cut off by the bath temperature rather than the bare magnon gap $\mu_0$, the scaling crosses over to $\s_0\propto T^{1/4}$. Finally, we investigate spin conductivity in the presence of electrical pumping. When the electrical pumping strength (denoted $\mu_s$) increases and approaches the critical value $\mu_s^c$ necessary for BEC instability, we find that the dc conductivity diverges as $\s_0\propto(\mu_s^c-\mu_s)^{-1/2}$. 

At the end of the work, motivated by the recent spin transport experiment on magnon gases close to the electrically-driven BEC,\cite{wimmerCM18} we discuss how the temperature scalings and the inverse square-root divergence of the spin conductivity can be verified using a similar experimental setup. 

The paper is organized as follows. In Sec.~\ref{model}, we introduce the models for the magnon system and the baths, and, in Sec.~\ref{theory}, obtain an effective theory for the magnons alone by tracing out all of the bath degrees of freedom using the Keldysh functional integral formalism. In Sec.~\ref{hft}, we develop a self-consistent Hartree-Fock theory for the magnon system, and study the BEC critical line and the finite temperature crossover behavior in Sec.~\ref{criticality}. Spin conductivity is computed in Sec.~\ref{sigma}, and a discussion on how the conductivity can be experimentally verified in a two-terminal spin transport setup is presented in Sec.~\ref{discussion}. Finally, conclusions are drawn in Sec.~\ref{conc}.

\section{Model}
\label{model}
\begin{figure}[t]
\includegraphics[width=\linewidth]{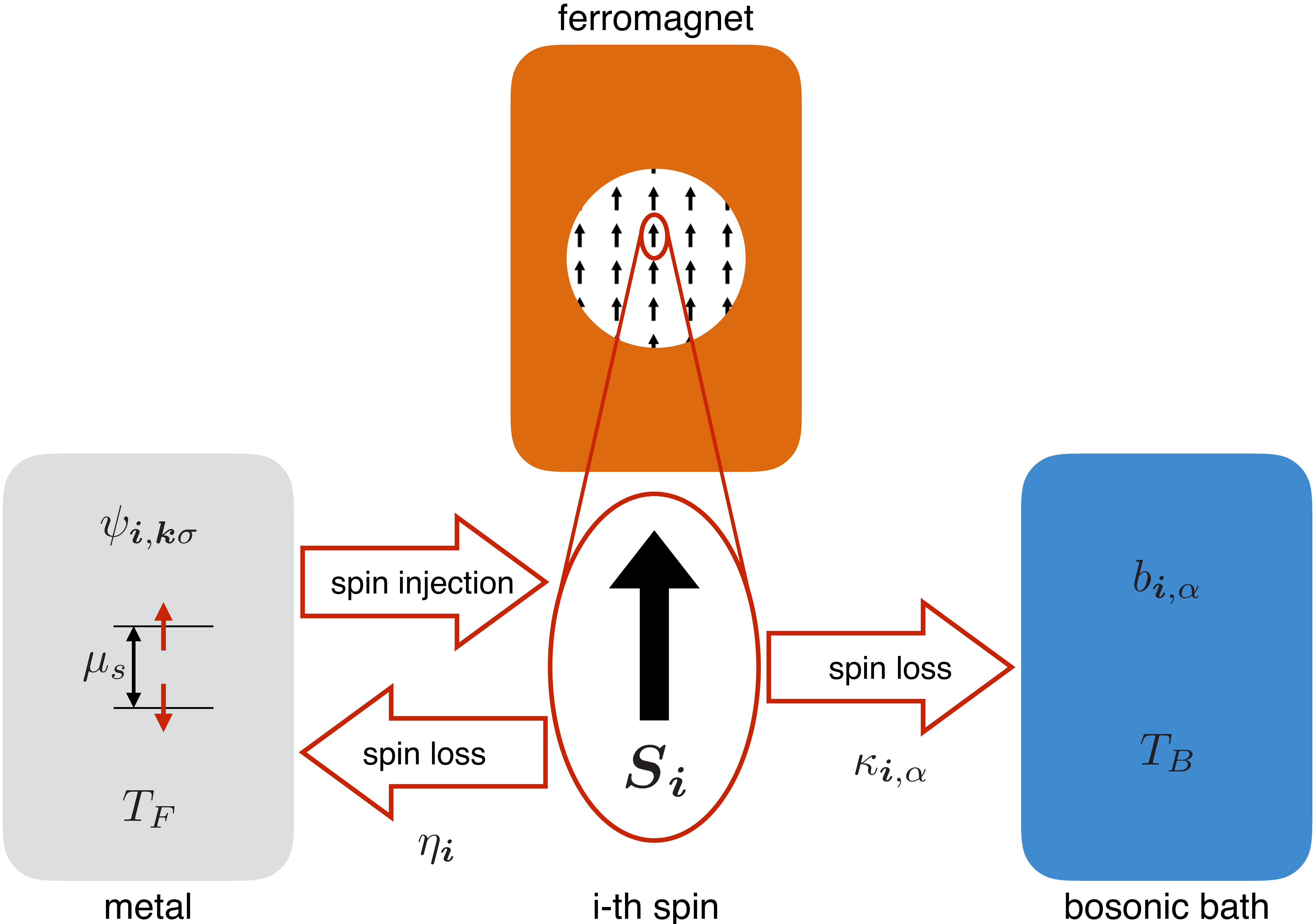}
\caption{(color online) Each atomic spin at site $\bi$ of the ferromagnetic insulator is coupled to its own phonon and metallic baths with coupling strengths $\eta_\bi$ and $\ka_{\bi,\al}$, respectively. The phonon bath is characterized by excitations $b_{\bi,\al}$, where $\al$ labels the eigenmodes, and the metallic bath is characterized by fermion excitations $\y_{\bi,\bk\s}$ with wavevector $\bk$ and spin $\s$. We assume the two baths are thermalized at temperatures $T_F$ and $T_B$, respectively, and a nonequilibrium spin accumulation $\mu_s$ in the metal allows for spin injection into the ferromagnet.}
\label{fig1}
\end{figure}

We consider a three-dimensional ferromagnetic insulator coupled to a bosonic bath and a metallic (fermionic) bath as shown in Fig.~\ref{fig1}. The boson bath is introduced to explicitly model the intrinsic spin/energy loss (i.e., Gilbert damping) in the ferromagnet, while the metal leads to additional Gilbert damping due to the fermionic continuum. A nonequilibrium spin accumulation $\mu_s$ in the metal (generated, e.g., via the spin Hall effect) also allows for incoherent spin injection into the magnetic insulator. 

\subsection{Ferromagnet}
Since our focus is on the impact of dissipation and nonequilibrium drive on a ferromagnetic insulator, we consider a relatively simple, but quite general, model Hamiltonian $H_F$ for an exchange ferromagnet with an easy-plane magnetic anisotropy perpendicular to the $z$ axis and in a uniform magnetic field $B$ along the same axis, i.e.,
\beq
\label{hf}
H_F=\frac{1}{2}\sum_{\bi\bj}\square{-J_{\bi\bj}\bS_\bi\cdot\bS_{\bj}+K_{\bi\bj}S_\bi^zS^z_\bj}+\hbar\g B\sum_\bi S^z_\bi\ .
\eeq
Here, $\bi,\bj$ label the sites of the lattice, $J_{\bi\bj}>0$ is the ferromagnetic exchange matrix, $K_{\bi\bj}>0$ is the anisotropy matrix, $\g$ is the gyromagnetic ratio and $S$ is the saturated spin moment per lattice site. We assume throughout that both $J_{\bi\bj}$ and $K_{\bi\bj}$ depend on the lattice positions $\bR_\bi$ and $\bR_\bj$ only through their difference, and that $J_{\bi\bj}=J_{\bj\bi}$ and $K_{\bi\bj}=K_{\bj\bi}$. Furthermore, assuming a spherically symmetric exchange matrix with a spatial range $\xi_J$, i.e., $J_{\bi\bj}\propto e^{-|\bR_\bi-\bR_\bj|^2/2\xi_J^2}$, we may approximate its Fourier transform as $J_\bq=\sum_\bj J_{\bi\bj}e^{-\ii\bq\cdot(\bR_\bi-\bR_\bj)}=J_\bzero e^{-q^2\xi_J^2/2}$, where $J_\bzero\equiv J_{\bq=\bzero}$ and $\ii=\sqrt{-1}$. A similar result can be applied to a spherically symmetric anisotropy constant $K_{\bi\bj}$ with a spatial range of $\xi_K$, i.e., $K_\bq\approx K_\bzero e^{-q^2\xi_K^2/2}$.

Following a standard boson mapping (see, e.g., Ref.~\onlinecite{batyevJETP85}), Eq.~\eqref{hf} can be re-expressed in terms of creation and annihilation operators, $a^\dag_\bq$ and $a_\bq$, for magnons with wavevector $\bq$, i.e.,
\begin{multline}
\label{hfhp}
H_F=\sum_{\bq}(\ve_\bq+\mu_0)a^\dag_\bq a_\bq\\
+\frac{1}{\sN}\sum_{\{\bq_n\}}V_{\bq_1\bq_3}a^\dag_{\bq_1}a^\dag_{\bq_2}a_{\bq_3}a_{\bq_4}\de_{\bq_1+\bq_2,\bq_3+\bq_4}\ ,
\end{multline}
where $\ve_\bq=S(J_{\bzero}-J_\bq)\approx(J_\bzero S\xi_J^2/2)q^2$, valid for $q\ll\xi_J^{-1}$, is the magnon spectrum, $\mu_0=\hbar\g B-SK_{\bzero}$ is the bare magnon gap (tunable using the external field), and $\sN$ is the total number of lattice sites in the ferromagnet. Under the boson mapping, the four-magnon vertex takes the form
\beq
\label{vqq}
V_{\bq_1\bq_3}=\frac{K_{\bq_1-\bq_3}-J_{\bq_1-\bq_3}}{2}+\la(J_{\bq_1}+J_{\bq_3})\ ,
\eeq
where $\la=S(1-\sqrt{1-1/2S})$. 

\subsection{Coupling to the baths}
We now define the spin-bath coupling. As shown in Fig.~\ref{fig1}, we assume that the atomic spin $\bS_\bi$ on each lattice site $\bi$ couples to its own {\em independent} boson and fermion baths. Then the Hamiltonian for the entire system (magnons and the baths included) can be written as $H=H_F+\sum_\bi\sH_\bi$, where $\sH_\bi$ is the local Hamiltonian describing the coupling of the atomic spin at site $\bi$ to the baths. This local Hamiltonian may be written as $\sH_\bi=\sH_\bi^B+\sH_\bi^m$, where $\sH_\bi^B$ and $\sH^m_\bi$ model the coupling to the boson and fermion baths, respectively. 

We begin by defining the local Hamiltonian corresponding to the bosonic bath, i.e.,
\beq
\sH^B_\bi=\sum_{\al}\hbar\W_{\al} b^{\dag}_{\bi,\al}b_{\bi,\al}+\hbar\sum_{\al}\square{\ka_{\bi,\al}a_\bi b^{\dag}_{\bi,\al}+h.c.}\ ,
\eeq
where $b_{\bi,\al}$ is the annihilation operator for a bath boson in the $\bi$-th bath and eigenmode $\al$, which couples to a magnon at the site with strength $\ka_{\bi,\al}$. The second term describes the decay of a magnon into the bath via transmutation into a superposition of bath boson modes. We assume that all of the bosonic baths are identical and that they are held at the same temperature $T_B$ so that the excitations occupy the states according to the Bose-Einstein distribution, i.e., $\langle b^\dag_{\bi,\al}b_{\bi,\al'}\rangle=n_\al\de_{\al\al'}$, where $n_\al=(e^{\hbar\W_{\al}/k_BT_B}-1)^{-1}$. We hereafter refer to the bosonic excitations as {\em phonons} and $T_B$ as the phonon temperature. In the absence of the metal (and, therefore, driving), the magnons thermalize to the Bose-Einstein distribution of the phonons.

The metallic bath contribution to the local Hamiltonian can be written as
\beq
\sH^m_\bi=\sum_{\bk,\s}\hbar\nu_{\bk}\y^\dag_{\bi,\bk\s}\y_{\bi,\bk\s}+\sum_{\bk,\bk'}\square{\eta_\bi\y^\dag_{\bi,\bk\up}\y_{\bi,\bk'\down}a_\bi+h.c.}\ ,
\eeq
where $\y_{\bi,\bk\s}$ annihilates a fermion in the $\bi$-th bath with wavevector $\bk$ and spin $\s$, $\hbar\nu_{\bk}=\hbar^2k^2/2m$ is the usual quadratic spectrum with effective mass $m$, and $\eta_\bi$ quantifies the hybridization between the magnon and the fermions at site $\bi$. This standard form for the magnon-electron hybridization has been considered in many other works.\cite{benderPRL12,benderPRB14,zhengPRB17} We assume here that all of the fermionic baths are identical and that they are held at a common temperature $T_F$ so that the excitations occupy the states according to the Fermi-Dirac distribution $\langle\y^\dag_{\bi,\bk\s}\y_{\bi,\bk'\s'}\rangle=f_{\bk\s}\de_{\s\s'}\de_{\bk\bk'}$, where $f_{\bk\s}=[e^{(\hbar\nu_{\bk}-\mu_\s)/k_BT_F}+1]^{-1}$. Here, $\mu_\s$ is the spin-dependent chemical potential that models the nonequilibrium spin accumulation $\mu_s$ inside the metal, i.e., $\mu_s=\mu_\up-\mu_\down$.

\section{Theory}
\label{theory}
Long after the couplings to the baths are turned on, the magnon system is expected to approach a steady state in which spin and energy injected by the pumping process are precisely balanced by the spin and energy lost through dissipation. In this nonequilibrium steady state, the magnons should generally be characterized by a non-thermal distribution function, level broadening, etc., and their behavior must be solved for in the presence of both drive and decay. To account for these nonequilibrium properties, we formulate the model from Sec.~\ref{model} using the real-time Keldysh path integral formalism.\cite{kamenevBOOK11} This formalism provides a systematic, straightforward way to trace out the bath degrees of freedom and obtain an effective theory for the magnons that takes the dissipative and nonequilibrium effects into account. The formal procedure presented in this section closely follows Ref.~\onlinecite{kamenevBOOK11}, so we refer the reader to this reference for any necessary additional details. 

\subsection{The Keldysh path integral formalism}
\begin{figure}[t]
\includegraphics[width=0.9\linewidth]{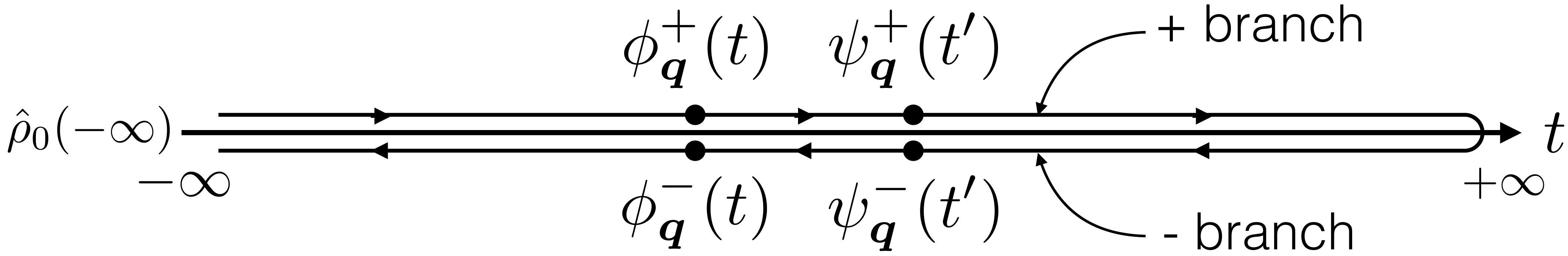}
\caption{Schwinger-Keldysh time-loop contour. The system is evolved with the full Hamiltonian $H$ from the infinite past $t=-\infty$ [at which point the density matrix is given by $\hat\rho_0(-\infty)$], to the infinite future $t=+\infty$, and then back to the infinite past. The forward evolution occurs along the $+$ branch and the backward evolution along the $-$ branch.}
\label{tl}
\end{figure}

Central to the Keldysh functional integral formalism is the Keldysh action $\cS$, which is obtained by time-evolving the system Lagrangian along the Schwinger-Keldysh time-loop contour (see Fig.~\ref{tl}). In deriving the nonequilibrium steady state properties, the formalism involves evolving the system with the full Hamiltonian from an initial state in the infinite past, at which point the density matrix of the total system factorizes into a product of equilibrium density matrices for the magnons and the baths, to the far future and back to the infinite past. To reduce the time integrals to unidirectional ones, however, it is customary to double the degrees of freedom at every point in time by defining separate fields on the forward and the backward branches (labeled by $+$ and $-$, respectively) of the time-loop contour (see Fig.~\ref{tl}). Following Ref.~\onlinecite{kamenevBOOK11}, we formulate $\cS$ in the $RAK$ basis, in which the Green functions on the time-loop contour are specified in terms of retarded (``$R$"), advanced (``$A$")and Keldysh (``$K$") components. In this basis, the bosonic fields on the forward and backward branches, $\f^+_\bq(t)$ and $\f^-_\bq(t)$, are rotated to the ``classical-quantum ($c$-$q$)" fields, $\f^c_\bq(t)$ and $\f^q_\bq(t)$, using
\beq
\f^{c,q}_\bq(t)=\frac{\f^+_\bq(t)\pm\f^-(t)}{\sqrt{2}}\ ,\ \ \ \bar\f^{c,q}_\bq(t)=\frac{\bar\f^+_\bq(t)\pm\bar\f^-(t)}{\sqrt{2}}\ ,
\eeq
the fermionic fields on the contour, $\y^+_\bk(t)$ and $\y^-_\bk(t)$, are rotated to the ``1-2" fields, $\y^1_\bk(t)$ and $\y^2_\bk(t)$, using
\beq
\y^{1,2}_\bq(t)=\frac{\y^+_\bq(t)\pm\y^-(t)}{\sqrt{2}}\ ,\ \ \ \bar\y^{1,2}_\bq(t)=\frac{\bar\y^+_\bq(t)\mp\bar\y^-(t)}{\sqrt{2}}\ ,
\eeq
and the causality structures of the bosonic ($\hD$) and fermionic ($\hG$) Green functions become
\beq\label{caus}\begin{aligned}
\hD=\mat{D^K}{D^R}{D^A}{0}\ ,&\qquad\hG=\mat{G^R}{G^K}{0}{G^A}\ ,
\end{aligned}\eeq
where $R,A,K$ label the retarded, advanced and Keldysh components as mentioned above.

Formulating the action for the model in Sec.~\ref{model} on the time-loop contour and transforming the fields to the $RAK$ basis, the Keldysh action for the magnon subsystem [corresponding to Eq.~\eqref{hfhp}] becomes
\begin{multline}
\label{sf}
\cS_F=\sum_{\bq}\int\frac{d\W}{2\p}A^\dag_\bq(\W)\mat{D^K_{0\bq}(\W)}{D^R_{0\bq}(\W)}{D^A_{0\bq}(\W)}{0}^{-1}A_\bq(\W)\\
-\frac{1}{\sN}\sum_{\{\bq_n\}}\int_{-\infty}^\infty dt\ V_{\bq_1\bq_3}\Big[a_{\bq_1}^{c*}(t)a_{\bq_2}^{c*}(t)a_{\bq_3}^{c}(t)a_{\bq_4}^q(t)\\
+a_{\bq_1}^{q*}(t)a_{\bq_2}^{q*}(t)a_{\bq_3}^{q}(t)a_{\bq_4}^{c}(t)+c.c.\Big]\de_{\bq_1+\bq_2,\bq_3+\bq_4}\ ,
\end{multline}
where $A^\dag_{\bq}(\W)=(a^{c*}_{\bq}(\W)\ a^{q*}_{\bq}(\W))$ collects the classical and quantum components of the magnon field, $a_{\bq}(t)=\int(d\W/2\p)a_{\bq}(\W)e^{-\ii\W t}$, and the components of the unperturbed magnon propagator matrix are given by 
\begin{align}
\label{d0r}
D^R_{0\bq}(\W)&=\frac{1}{\W-(\ve_\bq+\mu_0)/\hbar+\ii\de}=D^{A*}_{0\bq}(\W)\ ,\\
D^K_{0\bq}(\W)&=\coth\round{\frac{\hbar\W}{2k_BT}}[D^R_{0\bq}(\W)-D^A_{0\bq}(\W)]\ ,
\end{align}
where $T$ is the magnon temperature in the infinite past and $\de>0$ is the usual infinitesimal parameter.

The Keldysh action for the $\bi$-th phonon bath and its coupling to the magnons at site $\bi$ becomes
\begin{multline}
\label{sbi}
\cS^B_\bi=\sum_{\al}\int\frac{d\W}{2\p}B^\dag_{\bi,\al}(\W)\mat{d^K_\al(\W)}{d^R_\al(\W)}{d^A_\al(\W)}{0}^{-1}B_{\bi,\al}(\W)\\
-\sum_{\al}\int\frac{d\W}{2\p}\square{\ka_{\bi,\al}B^\dag_{\bi,\al}(\W)\hat\tau_xA_\bi(\W)+h.c.}\ ,
\end{multline}
where $B^\dag_{\bi,\al}(\W)=(b^{c*}_{\bi,\al}(\W)\ b^{q*}_{\bi,\al}(\W))$, once again, collects the classical and quantum components of the bath field and the bath Green functions read
\beq
\label{bbgf}
\begin{aligned}
d^R_\al(\W)&=\frac{1}{\W-\W_\al+\ii\de}=d^{A*}_\al(\W)\ ,\\
d^K_\al(\W)&=-2\p\ii\coth\round{\frac{\hbar\W}{2k_BT_B}}\de(\W-\W_\al)\ ;
\end{aligned}
\eeq
we use $\hat\tau_x$, $\hat\tau_y$ and $\hat\tau_z$ to denote Pauli matrices acting in Keldysh space. 

Finally, the Keldysh action describing the $\bi$-th metallic bath and its coupling to the magnons is given by
\begin{multline}
\label{smi}
\cS^m_\bi=\sum_{\bk,\s}\int\frac{d\w}{2\p}\Psi^\dag_{\bi,\bk\s}(\w)\mat{g^R_{\bk\s}(\w)}{g^K_{\bk\s}(\w)}{0}{g^A_{\bk\s}(\w)}^{-1}\Psi_{\bi,\bk\s}(\w)\\
-\frac{\eta_\bi}{\sqrt{2}}\sum_{\bk\bk'}\int\frac{d\w}{2\p}\frac{d\w'}{2\p}\Big[\Psi^\dag_{\bi,\bk\up}(\w)a^c_\bi(\w-\w')\Psi_{\bi,\bk\down}(\w')\\
+\Psi^\dag_{\bi,\bk\up}(\w)a^q_\bi(\w-\w')\hat\tau_x\Psi_{\bi,\bk\down}(\w')\Big]+h.c.\ ,
\end{multline}
where $\Psi^\dag_{\bi,\bk\s}(\w)=(\y^{1*}_{\bi,\bk\s}(\w)\ \y^{2*}_{\bi,\bk\s}(\w))$ collects the 1 and 2 components of the bath fermions, and the unperturbed bath fermion propagators are given by
\beq
\label{fbgf}
\begin{aligned}
g^R_{\bk}(\w)&=\frac{1}{\w-\nu_\bk+\ii\de}=g^{A*}_\bk(\w)\\
g^K_{\bk\s}(\w)&=-2\p\ii\tanh\round{\frac{\hbar\w-\mu_\s}{2k_BT_F}}\de(\w-\nu_\bk)\ .
\end{aligned}
\eeq
The full Keldysh partition function $Z$ can then be written as
\beq
\label{z}
Z=\int\frac{\sD\{A_\bq(\W),B_{\bi,\al}(\W),\Psi_{\bi,\bk\s}(\w)\}}{\tr\{\hat\rho_0(-\infty)\}}\exp\{\ii\cS\}\ ,
\eeq 
where $\cS=\cS_F+\sum_\bi(\cS_\bi^B+\cS_\bi^m)$ and $\hat\rho_0(-\infty)$ is the density matrix in the infinite past.

\subsection{Tracing out the baths}
\label{trace}
We may now integrate over the bath degrees of freedom $B_{\bi,\al}(\W)$ and $\Psi_{\bi,\bk\s}(\w)$ in Eq.~\eqref{z} in order to obtain an effective theory solely in terms of the magnons. The assumption here is that each bath is infinite and therefore remains unperturbed from its equilibrium configuration even in the presence of the (possibly strongly nonequilibrium) magnon system. The effects of the baths on the magnon subsystem can then be fully taken into account by performing a functional integral over the baths. 

We notice from Eqs.~\eqref{sbi} that the bosonic bath degrees of freedom can be integrated out using standard Gaussian integrals. Upon performing these integrals, we find (see Appendix~\ref{traceapp} for more details)
\beq
\label{sbi2}
\cS^B_\bi=-\int\frac{d\W}{2\p}A^\dag_\bi(\W)\mat{0}{\S^A_\bi(\W)}{\S^R_\bi(\W)}{\S^K_\bi(\W)}A_\bi(\W)\ ,
\eeq
where $\S^{R,A,K}_\bi(\W)$ is the phonon-induced magnon self-energy. As shown in Appendix~\ref{traceapp}, if we assume an ohmic form for the phonon bath spectral density, the retarded and advanced components of the self-energy reduce to 
\beq
\label{sigr}
\S^{R,A}_\bi(\W)=\mp\ii\al^B_\bi\W\ ,
\eeq
where $\al^B_\bi$ is a site-dependent Gilbert damping parameter, and the Keldysh component relates to the retarded and advanced components through the fluctuation-dissipation theorem,
\beq
\label{sigk}
\S^K_\bi(\W)=-2\ii\al_\bi^B\W\coth\round{\frac{\hbar\W}{2k_BT_B}}\ .
\eeq
We therefore see that an ohmic phonon bath leads to the standard level broadening proportional to the magnon frequency and thus to the familiar Gilbert damping phenomenology. 

As seen from Eq.~\eqref{smi}, the metallic bath degrees of freedom can also be integrated out using standard Gaussian integrals. However, tracing out these fermions is relatively more complex compared to the previous bosonic case because the fermions and the magnons couple nonlinearly. Therefore, once the Gaussian integrals over $\Psi_{\bi,\bk\s}(\w)$ are performed, corrections to the effective magnon action arise at all even orders in the magnon fields $A_\bi(\W)$ and renormalize the multi-magnon scattering vertices as well. 

At Gaussian order in the magnon fields, the correction renormalizes the Gaussian magnon action, i.e., the first term in Eq.~\eqref{sf},
\beq
\label{smi2}
\cS^{m(2)}_\bi=-\int\frac{d\W}{2\p}A^\dag_\bi(\W)\mat{0}{\Pi^A_\bi(\W)}{\Pi^R_\bi(\W)}{\Pi^K_\bi(\W)}A_\bi(\W)\ ,
\eeq
where the components of the fermion-induced magnon self-energy matrix are given by (see Appendix~\ref{tracefapp} for details)
\begin{align}
\label{pir}
\Pi^R_\bi(\W)&=-\ii\al^F_{\bi}(\W-\mu_s/\hbar)\\
\label{pik}
\Pi^K_\bi(\W)&=\coth\round{\frac{\hbar\W-\mu_s}{2k_BT_F}}\square{\Pi^R_\bi(\W)-\Pi^A_\bi(\W)}\ .
\end{align}
Here, $\al^F_{\bi}=\p|\eta_\bi|^2\rho_0^2\hbar^2$ is the site-dependent Gilbert damping arising from the metallic bath and $\rho_0$ is the bath fermion density of states at the Fermi level. 

The next order correction comes at fourth order in the magnon fields and renormalizes the quartic vertices in the magnon action [see the last two lines in Eq.~\eqref{sf}]. As shown in Appendix~\ref{tracefapp}, the tracing out of the fermion bath not only generates corrections to the existing quartic coefficients but also generates new quartic terms with even powers of quantum fields $a^q_\bq$. However, the bath only generates dissipative vertices | quartic terms with purely imaginary coefficients | that are $\mathcal{O}[(\al^F_\bi)^2]$, and, within the self-consistent Hartree-Fock approximation introduced below, they are expected to give subleading corrections to the dissipative effects already accounted for at the Gaussian order [see Eqs.~\eqref{pir} and \eqref{pik}]. We therefore ignore these metal-induced renormalizations of the quartic vertices in the remainder of the work.\footnote{Imaginary quartic coefficients for a dissipative Bose gas were obtained phenomenologically using a Markovian master equation approach in the context of exciton-polariton condensates.\cite{siebererRPP16} The current work provides a concrete microscopic derivation for how these dissipative quartic vertices can arise in the context of a driven, dissipative magnon gas (see Appendix~\ref{tracefapp} for more details).}

\section{Self-consistent Hartree-Fock Theory}
\label{hft}
We now discuss the properties of the driven, dissipative magnon subsystem {\em exclusively} in the normal (uncondensed) phase and discern the location of the BEC instability as a function of various system parameters such as the external magnetic field $B$ and the bath temperatures $T_B$ and $T_F$. This will be done by generalizing the self-consistent Hartree-Fock approach used for equilibrium dilute imperfect Bose gases\cite{fetterBOOK71} to the current nonequilibrium problem. For simplicity, we hereafter restrict our discussion to uniform magnon-bath coupling amplitudes, i.e., $\ka_{\bi,\al}\rightarrow\ka_\al$ and $\eta_\bi\rightarrow\eta$, though lifting this assumption is not expected to change the qualitative results presented below. 

The starting point for the discussion here is the effective magnon theory resulting from the elimination of the bath fields. Once these fields are traced out, the Keldysh partition function Eq.~\eqref{z} reduces to
\beq
Z=\int\frac{\sD\{A_\bq(\W)\}\exp\{\ii\bar\cS_F\}}{\tr\{\hat\rho_0^m(-\infty)\}}\ ,
\eeq 
where $\hat\rho_0^m(-\infty)$ is the density matrix for the magnons in the infinite past and the effective Keldysh action for the magnons reads
\begin{align}
\label{sfeff}
\bar\cS_F&=\cS_F+\sum_\bi\round{\cS_\bi^B+\cS_\bi^{m(2)}}\\
&=\sum_{\bq}\int\frac{d\W}{2\p}A^\dag_\bq(\W)\mat{D^K_{\bq}(\W)}{D^R_{\bq}(\W)}{D^A_{\bq}(\W)}{0}^{-1}A_\bq(\W)\\
&\quad-\frac{1}{\sN}\sum_{\{\bq_n\}}\int_{-\infty}^\infty dt\ V_{\bq_1\bq_3}\Big[a_{\bq_1}^{c*}(t)a_{\bq_2}^{c*}(t)a_{\bq_3}^{c}(t)a_{\bq_4}^q(t)\\
&\quad+a_{\bq_1}^{q*}(t)a_{\bq_2}^{q*}(t)a_{\bq_3}^{q}(t)a_{\bq_4}^{c}(t)+c.c.\Big]\de_{\bq_1+\bq_2,\bq_3+\bq_4}\ .
\end{align}
We note that the magnon propagator matrix above [with the components $D^{R,A,K}_\bq(\W)$] now fully accounts for the nonequilibrium drive and dissipation due to the baths. Using Eqs.~\eqref{d0r}, \eqref{sigr}, and \eqref{pir}, its retarded component (which relates to the advanced component by complex conjugation) can easily be read off from Eq.~\eqref{sfeff}, 
\beq
\label{dr}
D^R_\bq(\W)=\frac{1}{\W-(\ve_\bq+\mu_0)/\hbar+\ii\al_B\W+\ii\al_F(\W-\mu_s/\hbar)}\ ,
\eeq
where $\al_B$ and $\al_F$ can now be interpreted as the (spatially uniform) Gilbert damping parameters arising from the phonon and metallic baths, respectively. 

Using Eqs.~\eqref{sf}, \eqref{sbi2}, and \eqref{smi2}, the Keldysh component of the magnon propagator can also be read off directly from Eq.~\eqref{sfeff} as
\beq
\label{dkgen}
D_\bq^K(\W)=\frac{\ii}{2}\frac{\Pi^K(\W)+\Sigma^K(\W)}{\al_B\W+\al_F(\W-\mu_s/\hbar)}[D^R_\bq(\W)-D^A_\bq(\W)]\ .
\eeq
However, noting that the Keldysh propagator can generally be related to the retarded and advanced components via the magnon distribution function $N(\W)$, i.e., $D^K_\bq(\W)=[2N(\W)+1][D^R_\bq(\W)-D^A_\bq(\W)]$,\cite{kamenevBOOK11} Eq.~\eqref{dkgen} together with Eqs.~\eqref{sigk} and \eqref{pik} allow us to directly extract the nonequilibrium magnon distribution function,
\begin{multline}
\label{n}
N(\W)=\frac{1}{\al\W-\al_F\mu_s/\hbar}\\
\times\square{\frac{\al_B\W}{e^{\hbar\W/k_BT_B}-1}+\frac{\al_F(\W-\mu_s/\hbar)}{e^{(\hbar\W-\mu_s)/k_BT_F}-1}}\ ,
\end{multline}
where $\al=\al_B+\al_F$ is the total Gilbert damping parameter. 

We see from Eq.~\eqref{n} that the magnon distribution function, in general, has a nonthermal form in the presence of mismatched bath temperatures and/or electrical pumping. Even at zero pumping (i.e., $\mu_s=0$), $N(\W)$ is given by a linear combination of two Bose-Einstein distribution functions, one corresponding to the phonons equilibrated at temperature $T_B$ and the other corresponding to the spin-1 particle-hole excitations in the metal equilibrated at temperature $T_F$, and they are weighted by the respective Gilbert damping parameters associated with each of the baths. This has important consequences on the finite temperature crossover behavior for the magnons close to the BEC critical line, as we later show in Sec.~\ref{finitet}. 

The steady-state non-thermal magnon distribution function obtained here contrasts with similar past studies of electrically pumped magnon BECs, in which magnons were assumed to be internally thermalized to the Bose-Einstein distribution function with a well-defined effective temperature and a chemical potential.\cite{benderPRB14,fjaerbuPRB17}
\begin{figure}[t]
\includegraphics[width=0.8\linewidth]{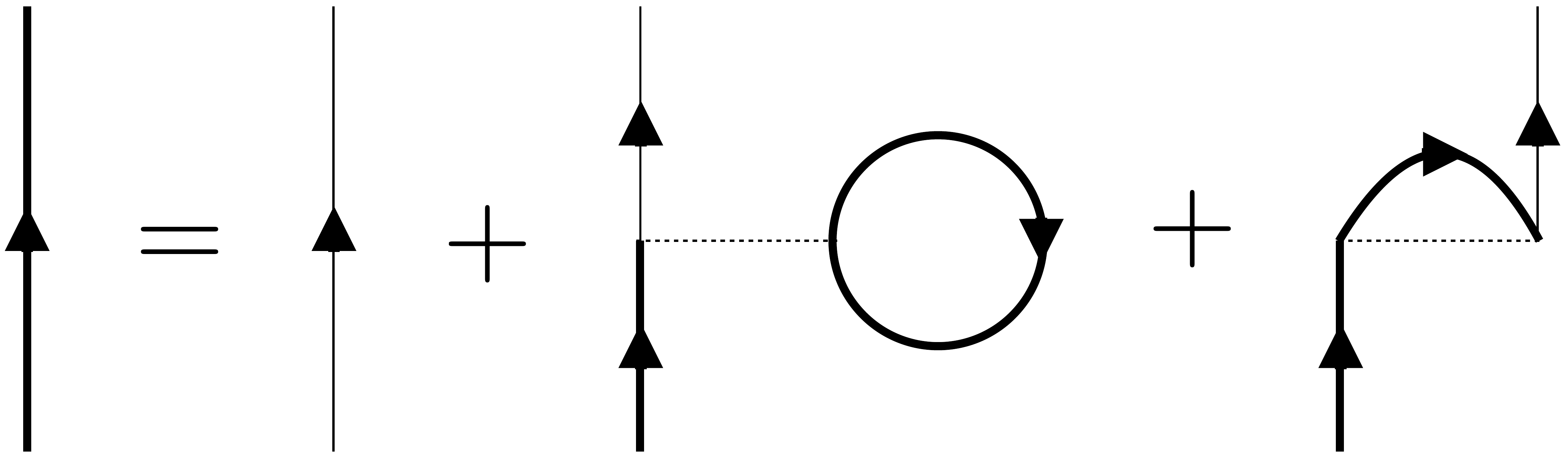}
\caption{Dyson equation for the full Green function in the self-consistent Hartree-Fock theory. Thin lines denote noninteracting magnon propagators and heavy lines the full propagators. The dotted lines correspond to the interaction vertex $V_{\bq\bq'}$.}
\label{fig2}
\end{figure}

\subsection{Self-consistent conditions}
The Dyson equation for the full magnon Green function within the self-consistent Hartree-Fock approximation is diagrammatically shown in Fig.~\ref{fig2}, where each magnon propagator has the bosonic Keldysh matrix structure consistent with Eq.~\eqref{caus}. Each thin line represents the Gaussian magnon propagator matrix, i.e., $D^{R,A,K}_\bq(\W)$, while the thick lines correspond to the full propagator matrix. Within the self-consistent theory, the components of the Keldysh one-loop self-energy matrix are given by
\beq
\tSig^R_\bq=-\frac{4}{\sN}\frac{1}{\hbar}\sum_{\bq'}\int\frac{d\W}{2\p}N(\W)\round{V_{\bq\bq'}+V_{\bq'\bq'}}{\rm Im}\sD^R_{\bq'}(\W)=\tSig^A_\bq\ ,
\eeq
where $\sD^{R}_\bq(\W)$ is the retarded component of the full self-consistent magnon propagator, and we find $\tSig^K_\bq=0$. This then leads to a self-consistent magnon spectrum given by
\begin{multline}
\label{sceq}
\tilde\ve_\bq+\mu=\ve_\bq+\mu_0\\
-\frac{4}{\sN}\sum_{\bq'}\int\frac{d\W}{2\p}N(\W)\round{V_{\bq\bq'}+V_{\bq'\bq'}}{\rm Im}\{\sD^R_{\bq'}(\W)\}\ ,
\end{multline}
where the full retarded magnon propagator above is defined with the renormalized spectrum $\tilde\ve_\bq$ and band gap $\mu$, i.e.,
\beq
\label{Dr}
\sD^R_\bq(\W)=\frac{1}{\W-(\tilde\ve_\bq+\mu)/\hbar+\ii\al\W-\ii\al_F\mu_s/\hbar}\ .
\eeq
For weak damping, i.e., $\al_B,\al_F\ll1$, the magnon spectral function, $\cB_\bq(\W)\equiv-2{\rm Im}\sD^R_{\bq}(\W)$, is strongly peaked near $q\sim[2(\hbar\W-\mu)/J_\bzero S\xi^2_J]^{1/2}$. Therefore, contributions to the $\bq'$-summation coming from large wavevectors $q'\sim\xi^{-1}_J$ in Eq.~\eqref{sceq} (assuming that the spatial range of the exchange constant is on the order of a few lattice constants) is exponentially suppressed by $N(\W)$ as long as $J_\bzero S/k_BT_B, J_\bzero S/k_BT_F\gg 1$, which is almost always true. The last term in Eq.~\eqref{sceq} can then be approximated well by setting $\bq'=0$ in $V_{\bq\bq'}$ and $V_{\bq'\bq'}$. Using Eq.~\eqref{vqq} and recalling that $J_\bq\approx J_\bzero e^{-q^2\xi_J^2/2}$ and $K_\bq\approx K_\bzero e^{-q^2\xi_K^2/2}$, we may then expand $V_{\bq\bzero}$ to quadratic order in $q\xi_J\ll1$ and $q\xi_K\ll1$ and obtain the self-consistent equation as
\begin{multline}
\label{sceq2}
\tilde\ve_\bq+\mu=\ve_\bq+\mu_0+2n[(K_\bzero+J_\bzero(4\la-1)]\\
-\frac{K_\bzero(q\xi_K)^2+J_\bzero(q\xi_J)^2(2\la-1)}{2}n\ ,
\end{multline}
where 
\beq
n=-\frac{2}{\sN}\sum_{\bq'}\int\frac{d\W}{2\p}N(\W){\rm Im}\{\sD^R_{\bq'}(\W)\}
\eeq
is the magnon number per lattice site. 

The last term on the right hand side of Eq.~\eqref{sceq2} leads to the renormalization of the effective magnon mass. However, this renormalization remains small in the dilute magnon limit, i.e., $n/S\ll1$. We ignore this correction and finally arrive at the self-consistent equation for the magnon gap,
\beq
\label{mass}
\mu=\mu_0+2n[K_\bzero+J_\bzero(4\la-1)]\ .
\eeq
This self-consistent equation for the magnon gap encodes the nonequilibrium effects (i.e., mismatch in the bath temperatures and electrical pumping) through the magnon density $n$ in the second term. Solving this equation therefore allows us to study the location of the BEC instability and the various crossover behavior near criticality for the general nonequilibrium magnon system. A similar self-consistent equation was obtained recently using the Keldysh formalism for a spin-current driven magnetic insulator in Ref.~\onlinecite{troncosoPRB19}.

\section{BEC Critical Phenomena}
\label{criticality}
We now construct a two-dimensional phase diagram for the magnon system spanned by the phonon temperature $T_B$ and external magnetic field $B$. We define the equilibrium limit as $T_B=T_F\equiv T$ and $\mu_s=0$, and systematically study the departures from this limit by introducing $T_F\ne T_B$ and $\mu_s>0$. 

We begin with the limit of zero electrical pumping, i.e., $\mu_s=0$. In this limit, the BEC critical line [defined by $\mu=0$ in Eq.~\eqref{mass}] is given by the following critical magnetic field
\beq
\label{bc}
\hbar\g B_c=SK_\bzero-\G_\al c\square{\frac{\al_B}{\al}T_B^{3/2}+\frac{\al_F}{\al}T_F^{3/2}}\ ,
\eeq
where $c$ is a real constant given by
\beq
\label{cbcf}
c=[K_\bzero+J_\bzero(4\la-1)]\frac{v_s\z_{3/2}(1)}{\sqrt{2}\xi_J^3}\round{\frac{k_B}{\p J_\bzero S}}^{3/2}\ ,
\eeq
$\z_{3/2}(x)$ is the polylog function, $v_s$ is the volume occupied by each atomic spin, and $\G_\al=(1+\al^2)^{1/4}\cos[(\tan^{-1}\al)/2]$ is a prefactor that arises due to dissipation. 

\begin{figure}[t]
\includegraphics[width=0.85\linewidth]{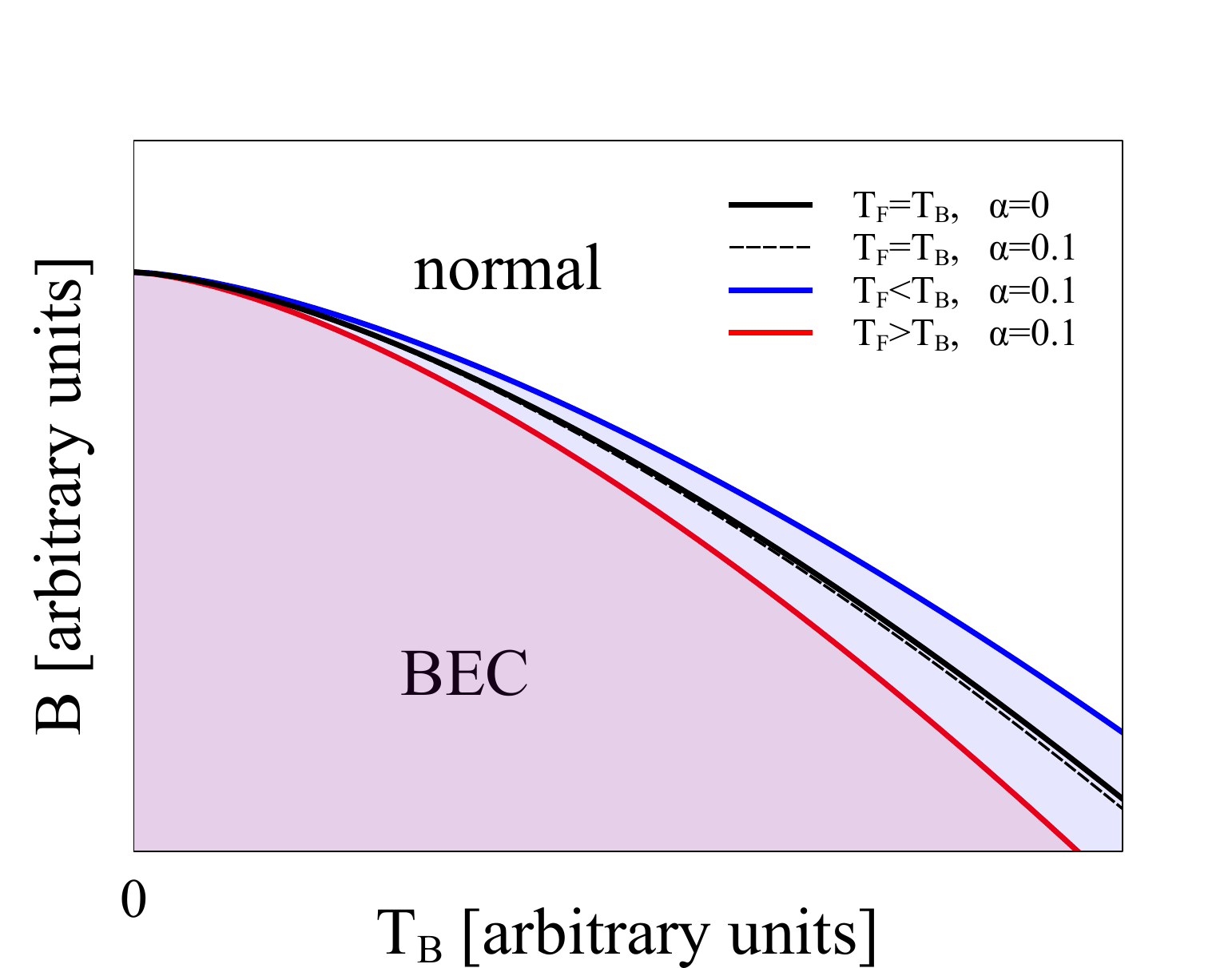}
\caption{(color online) Critical lines separating the BEC phase from the normal phase at zero electrical pumping, $\mu_s=0$. The phase diagram is spanned by the magnetic field $B$ and phonon temperature $T_B$. The solid black line corresponds to the equilibrium, isolated limit with $T_B=T_F$ and $\al=0$. For $\al=0.1$, the dashed line obtains, indicating that the BEC region is shrunk by dissipation. The red (blue) line corresponds to the critical line with $T_F=1.2T_B$ ($T_B=0.8T_F$); here, we have chosen $\al_B=\al_F$.}
\label{fig3}
\end{figure}

If the two bath temperatures are equal and the dissipation is absent, i.e., $T_B=T_F$ and $\al=0$, we have $\G_{\al=0}=1$ and Eq.~\eqref{bc} reduces precisely to the well-known BEC critical line for dilute Bose gases (see solid black line in Fig.~\ref{fig3}).\cite{fetterBOOK71,sachdevBOOK11} If we then introduce Gilbert damping $\al>0$ (while still keeping $T_F=T_B$), the critical line is modifed to the dashed line; we find that dissipation, as expected, leads to the shrinking of the BEC phase region.

Let us now consider both $T_F\ne T_B$ and $\al>0$. The two colored lines in Fig.~\ref{fig3} give the critical lines for mismatched bath temperatures, $T_F\ne T_B$, the red and blue lines corresponding to $T_F>T_B$ and $T_F<T_B$, respectively. We find that raising (lowering) the fermion bath temperature above (below) the phonon temperature generally destabilizes (stabilizes) the BEC phase. In the absence of electrical pumping $\mu_s=0$, our model therefore predicts that triggering a steady state magnon BEC with a {\em positive} temperature difference between the metal and the phonons ($T_F-T_B>0$, i.e., magnon BEC via spin Seebeck mechanism alone) is not possible.

The reason for this claim can be understood as follows. Our theoretical analysis in Sec.~\ref{hft} shows that once the phonon and metallic baths are integrated out, magnon distribution function Eq.~\eqref{n} is no longer thermal but is defined by the distribution functions of both baths. The magnon distribution function therefore ``floats" (i.e., adjusts) according to the thermal distribution functions of the surrounding baths. In this sense, it would be reasonable to expect that realizing magnon BEC by heating the metal is impossible because that will always heat up the magnon system and hinder condensation. 

Electrically-pumped BEC of magnons in a setup very similar to Fig.~\ref{fig1} was studied recently in Ref.~\onlinecite{benderPRB14} by assuming that magnons remain internally thermalized with an effective temperature and chemical potential. In the so-called ``floating temperature" regime, the work obtains the effective magnon temperature by balancing the spin/energy transfer across the metal-magnet interface and loss to the  phonon bath. In this regime, the work shows that raising (lowering) the metal temperature above (below) the phonon temperature tends to hinder (facilitate) magnon condensation, and therefore shows some level of consistency with our findings.


\subsection{Finite temperature crossovers}
\label{finitet}
The finite temperature crossover behavior of the standard equilibrium Bose gas near BEC instability has been studied in detail in, e.g., Ref.~\onlinecite{sachdevBOOK11}. However, we show in this section that the presence of two independent bath temperatures endows the current system with a richer thermal crossover behavior than the standard counterpart. Here, we start by maintaining $\mu_s=0$ and study the effects of the two bath temperatures in the vicinity of the $T_B=T_F=0$ quantum critical point located at $\mu_0=0$. 

\begin{figure}[t]
\includegraphics[width=0.8\linewidth]{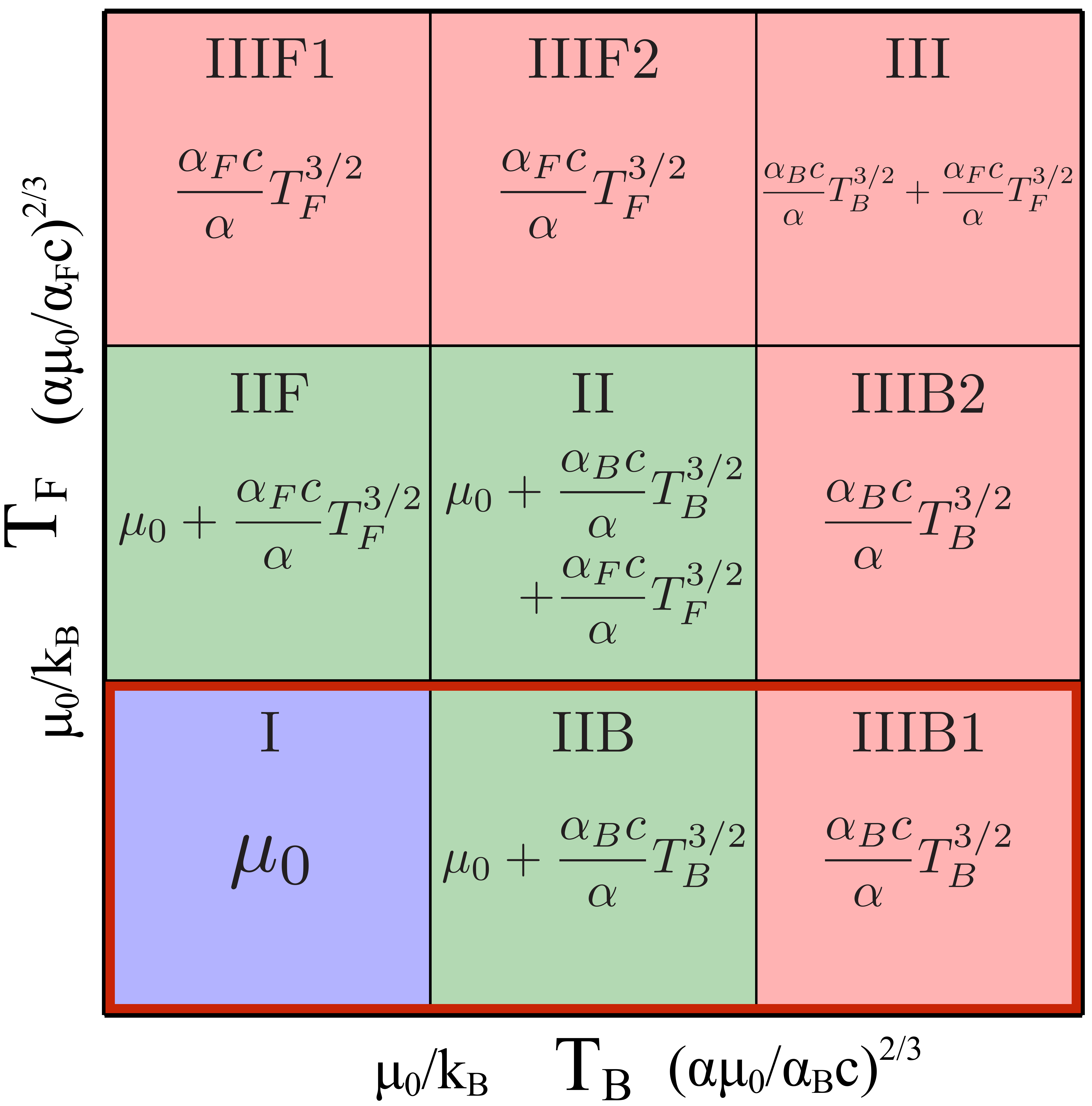}
\caption{(color online) Approximate expressions for the magnon gap $\mu$ for $\mu_s=0$. The columns separate the low, intermediate and high temperature regimes with respect to the phonon temperature, while the rows separate the three regimes with respect to the metal temperature. The three finite-temperature regimes familiar from the equilibrium dilute Bose gas\cite{sachdevBOOK11} are enclosed by the thick red box.}
\label{fig4}
\end{figure}

For weak dissipation, $\al_B,\al_F\ll1$, we find that the effects of the baths on the magnon density of states $\rho_m(\W)$ are very weak, so we approximate $\rho_m$ with its expression in the isolated limit, i.e., 
\beq
\label{dos}
\rho_m(\W)=-\left.\frac{1}{\hbar}\sum_\bq{\rm Im}\sD^R_\bq(\W)\right|_{\substack{\al_B=0\\\al_F=0}}\approx\frac{\sV\sqrt{\hbar\W-\mu}\thi(\hbar\W-\mu)}{\sqrt{2}\p(J_\bzero S\xi_J^2)^{3/2}}\ ,
\eeq
where $\sV$ is the volume of the ferromagnet. Then the self-consistent condition Eq.~\eqref{mass} for the magnon gap becomes
\begin{multline}
\label{scmu}
\mu=\mu_0+\frac{\al_B}{\al}\square{\frac{\z_{3/2}(e^{-\mu/k_BT_B})}{\z_{3/2}(1)}}cT_B^{3/2}\\
+\frac{\al_F}{\al}\square{\frac{\z_{3/2}(e^{-\mu/k_BT_F})}{\z_{3/2}(1)}}cT_F^{3/2}\ .
\end{multline}
This equation can be solved for $\mu$ by iteration and gives rise to nine different regimes with respect to the two bath temperatures as shown in Fig.~\ref{fig4}. Since the gap $\mu$ is also directly related to the correlation length $\xi$ of the superfluid order parameter fluctuations via $\xi^{-2}\propto\mu$, Fig.~\ref{fig4} also summarizes how the correlation length of the critical fluctuations, which diverges right at the critical line, is cutoff by departures from the critical line.

Let us first touch base with the known past results. The finite temperature crossover behavior of the correlation length for the standard equilibrium Bose gas\cite{sachdevBOOK11} can be reproduced by setting $\al_F=0$ here and varying the phonon temperature $T_B$. If $\al_F=0$, magnons thermalize to the phonons, so $T_B$ here defines the magnon temperature. The three known regimes that arise are enclosed in the thick red box in Fig.~\ref{fig4}. At low temperatures $T_B\ll\mu_0/k_B$ (shaded in blue), the correlation length is essentially cutoff by the bare gap $\mu_0$ and its $T_B$-dependent corrections are exponentially small. In this regime, the inter-particle spacing between the magnons is much larger than the thermal de Broglie wavelength and the the magnon density is exponentially small. In the intermediate temperature regime $\mu_0/k_B\ll T_B\ll(\mu_0/c)^{2/3}$ (shaded in green), $\xi^{-2}$ is still dominated by $\mu_0$, however, the subleading correction now possesses a power-law form in contrast to the exponential form obtained in the low-temperature regime. In this regime, the magnon density scales with temperature as $T_B^{3/2}$ and the inter-particle spacing becomes of order the de Broglie wavelength. Finally, in the high-temperature regime $T_B\gg(\mu_0/c)^{2/3}$ (shaded in red), the correlation length is cutoff by temperature and one finds $\xi\sim T_B^{-3/4}$.

If we now re-introduce the fermionic bath (i.e., $\al_F>0$), each of the above-mentioned regimes is further subdivided into three sub-regimes depending on the magnitude of the metal temperature $T_F$ (see Fig.~\ref{fig4}): the bottom row corresponds to the low temperature regime with respect to the fermion temperature with $T_F\ll\mu_0/k_B$, the middle row to intermediate fermion temperatures $\mu_0/k_B\ll T_F\ll(\al\mu_0/\al_Fc)^{2/3}$ and the top row to the high temperature regime $T_F\gg(\al\mu_0/\al_Fc)^{2/3}$. These new sub-regimes arise because the metal is just as capable of creating thermal magnons in the ferromagnet and inducing thermal crossovers in the correlation length as the phonon bath. We see that the correction to the magnon gap coming from $T_F$ is exponentially small in the low temperature regime and so $T_F$ essentially does not enter the expression for the correlation length (see the bottom row). However, in analogy with $T_B$, the metal temperature becomes increasingly effective in cutting off the correlation length as it increases, and once the high temperature regime is reached (see the top row), the correlation length essentially becomes defined by $T_F$ unless the phonon temperature $T_B$ is also in the high temperature regime (i.e., the top right corner).

\subsection{BEC instability due to electrical pumping}
\label{scgap}
We finally come to the discussion of the nonequilibrium drive and reinstate electrical pumping $\mu_s$. As one increases $\mu_s$ from zero, the instability of the normal phase is triggered once the frequency $\mu/\hbar$ corresponding to the excitations at the bottom of the band obeys ${\rm Im}\{\sD^{R-1}_\bq(\mu/\hbar)\}=0$, i.e., 
\beq
\label{inst}
\mu_s^c=\round{1+\frac{\al_B}{\al_F}}\mu\ .
\eeq 
If $\mu_s$ is increased beyond this point, the imaginary parts of the small-$q$ poles become positive and fluctuations in the system grow in time, thus signaling an instability.\footnote{In the absence of the phonon bath, i.e., $\al_B=0$, we see from Eq.~(\ref{inst}) that the instability occurs precisely when the nonequilibrium spin accumulation reaches the magnon gap (i.e., $\mu^c_s=\mu$). However, this threshold is raised in the presence of the phonons. This is reasonable because in the latter case, the drive must overcome additional decay to the phonons in order for the magnons to reach the critical density necessary for condensation. In Ref.~\onlinecite{benderPRB14}, the condition for instability toward magnon condensation was given by \beq\mu_s^c=\round{1+\frac{\al_B}{2\al_F}}\mu\ ,\eeq which differs from Eq.~(\ref{inst}) by a factor of 2 in front of $\al_F$. The factor arises in Ref.~\onlinecite{benderPRB14} because spin injection occurs at the metal-magnet boundary, which coincides with the antinodes of the thermal magnon normal modes. As evident from Sec.~\ref{model}, open boundaries are absent in the current work, so the factor of 2 does not arise here.} From Eq.~\eqref{Dr}, we see that the instability condition Eq.~\eqref{n} signals a divergence in the magnon distribution function $N(\W)$ as well.

We note that Eq.~\eqref{inst} is actually a self-consistent equation for $\mu_s$ because $\mu$ itself depends on $\mu_s$ through magnon density $n$ in Eq.~\eqref{mass}. However, we find that $\mu_s$ has a relatively small effect on the solution for the magnon gap $\mu$ (see Appendix~\ref{neqscmu} for more details). In other words, even with $\mu_s>0$, $\mu$ can be well-approximated by the solution of Eq.~\eqref{mass} with $\mu_s$ set to zero, and $\mu_s$ can essentially be introduced as an independent nonequilibrium parameter. The BEC instability criterion Eq.~\eqref{inst} can therefore be well-approximated by
\beq
\label{inst2}
\mu_s^c\approx\round{1+\frac{\al_B}{\al_F}}\mu(\mu_s=0,T_B,T_F)\ .
\eeq 

\section{Spin Conductivity}
\label{sigma}
Finite temperature crossovers exhibited by the magnon gap entails corresponding crossovers in the magnon spin conductivity. Here, we use the Kubo formalism to compute the magnon spin conductivity within the normal phase and the Hartree-Fock approximation for various bath temperatures $T_B$ and $T_F$ and the magnetic field $\mu_0$. We then later investigate the effects of finite electrical pumping $\mu_s>0$.

Spin conductivity can be obtained by looking at the response of the magnons to an external magnetic field $B(\bx,t)$; here, we focus only on the spin current polarized along the $z$ axis, which is the conserved component. The relevant (time-dependent) perturbation can then be written as
\beq
\label{hpb}
H'(t)=\hbar\g\int d^3\bx\, B(\bx,t)a^\dag(\bx)a(\bx)\ ,
\eeq
where $a(\bx)=a_\bi/v_s^{1/2}$ is the continuum magnon field operator. Assuming that the magnetic field has a negative uniform gradient only along the $x$ axis (see bottom of Fig.~\ref{fig6}), the resulting spin current $J(\bx,t)$ flowing along the same axis, within linear-response, reads~\cite{mahanBOOK90,alvarezPRB02,meierPRL03}
\beq
\label{jxt}
J(\bx,t)=\hbar\g\int d^3\bx'\int dt' \chi(\bx-\bx',t-t')B(\bx',t')\ ,
\eeq
where the susceptibility is defined as
\beq
\label{susc}
\chi(\bx-\bx',t-t')=-\frac{\ii}{\hbar}\thi(t-t')\langle[j(\bx,t),\varrho(\bx',t')]\rangle_{\rm HF}\ ,
\eeq
$\varrho(\bx,t)=a^\dag(\bx,t)a(\bx,t)$ is the magnon density operator and $j(\bx,t)=\ii(J_\bzero S\xi_J^2/2)[\pd_x a^\dag(\bx)]a(\bx)+h.c.$ is the magnon spin current operator. 

The spin conductivity $\s$ is defined as the constant of proportionality between the negative gradient of the magnetic field and the spin current that flows in response to it, i.e., $J_\bq(\w)=\s(\bq,\w)(-\ii q_x)\hbar\g B_\bq(\w)$, where $\s(\bq,\w)=(-\ii q_x)^{-1}\chi(\bq,\w)$. It can be calculated by closely following the standard calculation for the Kubo electrical conductivity in metals.\cite{mahanBOOK90} Moving the details of the calculations to Appendix~\ref{kubo}, the dc spin conductivity, within the Hartree-Fock approximation and the spatially uniform limit, becomes
\beq
\label{s0}
\s_0=-\frac{2}{3}\round{\frac{J_\bzero S\xi_J^2}{2\hbar}}^2\int\frac{d\W}{2\p}\int\frac{d^3\bk}{(2\p)^3}\pd_\W N(\W)k^2\cB^2_\bk(\W)\ .
\eeq

\subsection{Finite-temperature crossovers}
To explore the behavior of $\s_0$, we begin by setting $\mu_s=0$ and, for simplicity, consider two equal bath temperatures, i.e., $T_B=T_F\equiv T$. The dc spin conductivity can then be studied in the three temperature regimes, I, II and III, shown in Fig.~\ref{fig4}. Both the phonon and metallic baths contribute to the thermal population of magnons inside the ferromagnet. Evaluating Eq.~\eqref{s0} in the limit of weak damping $\al_B,\al_F\ll1$, $\s_0$ can be well approximated by
\beq
\s_{0}\approx\frac{1}{24\p^2\al}\sqrt{\frac{2k_BT}{J_\bzero S\xi_J^2}}\mathfrak{S}\round{\frac{\mu}{k_BT}}\ ,
\eeq
where the dimensionless function $\mathfrak{S}(s)$ has the following asymptotic behavior
\beq
\mathfrak{S}(s)\approx\left\{\begin{array}{lcl}\frac{3\sqrt{\p}}{4}\frac{e^{-s}}{s}&,&s\gg1\\\frac{3\p}{2}\frac{1}{\sqrt{s}}&,&s\ll1\end{array}\right.\ .
\eeq

In the low temperature regime $T\ll\mu_0/k_B$, the baths generate very few thermal magnons in the ferromagnet and the conductivity becomes exponentially suppressed. In this limit, we have $\mu\approx\mu_0$ and obtain
\beq
\label{sclow}
\s_{0}\approx\frac{\sqrt{2}}{32\al\xi_J}\round{\frac{J_\bzero S}{\mu_0}}\round{\frac{k_BT}{\p J_\bzero S}}^{3/2}e^{-\mu_0/k_BT}\ .
\eeq

When $T$ is in the intermediate to high temperature regimes $T\gg\mu_0/k_B$, the dimensionless function $\mathfrak{S}(s)$ crosses over to the other asymptotic regime, and we obtain
\beq
\label{scint}
\s_{0}\approx\frac{\sqrt{2}}{16\al\xi_J}\sqrt{\frac{J_\bzero S}{\mu_0+cT^{3/2}}}\round{\frac{k_BT}{\p J_\bzero S}}\ .
\eeq
In the intermediate temperature regime (i.e., regime II), the bare magnon gap still obeys $\mu_0\gg cT^{3/2}$ and the dc spin conductivity scales as $\s_0\sim T$. However, once the temperature is increased into the high temperature regime (i.e., regime III), the inequality reverses, i.e., $\mu_0\ll cT^{3/2}$, and the temperature scaling of the dc spin conductivity crosses over to $\s_0\sim T^{1/4}$. Experimental consequences of these results will be discussed in Sec.~\ref{discussion}.

\subsection{BEC instability due to electrical pumping}
\label{scdrive}
We now consider finite electrical pumping $\mu_s>0$ for a fixed set of bath temperatures $T_B$ and $T_F$. In Sec.~\ref{scgap}, we found that the magnon gap $\mu$ is essentially determined by $T_B$ and $T_F$ alone, and so $\mu_s$ can be thought of as an independent parameter. Exploiting this fact, we explore the behavior of $\s_0$ as one tunes $\mu_s$ toward the BEC instability point $\mu^c_s$ [see Eq.~\eqref{inst}] and find that $\s_0$ diverges algebraically. The relevant exponent can be found straightforwardly by evaluating Eq.~\eqref{s0} with $\mu_s\ne0$. We fix the magnon gap to a finite positive value $\mu>0$ and sweep $\mu_s$ toward $\mu_s^c$; we also assume that the bath temperatures do not change as electrical pumping is increased to the instability point. Then, again in the limit of weak damping $\al_{B},\al_F\ll1$, we find
\beq
\label{s0bias}
\s_0\approx \frac{1}{6\p^2}\sqrt{\frac{2}{J_\bzero S\xi_J^2}}\int_{\mu/\hbar}^\infty d\W\ N(\W)\frac{\pd}{\pd\W}\frac{(\hbar\W-\mu)^{3/2}}{\al\hbar\W-\al_F\mu_s}\ ,
\eeq
which diverge as $(\mu_s^c-\mu_s)^{-1/2}$ as $\mu_s\rightarrow\mu_s^c$. 

\section{Discussion}
\label{discussion}

\begin{figure}[t]
\includegraphics[width=\linewidth]{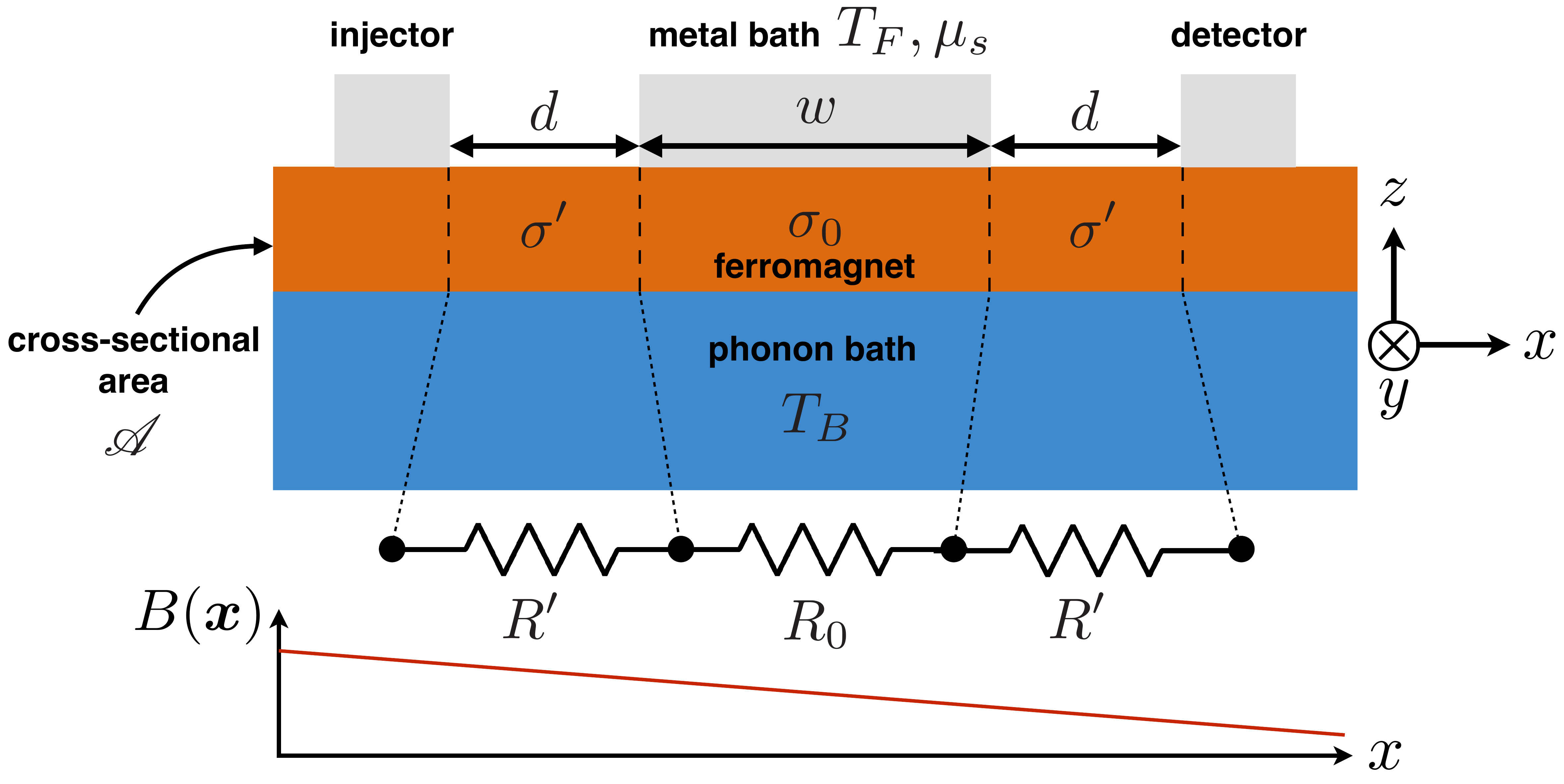}
\caption{(color online) A possible device for measuring spin conductivity. Magnons are injected into the ferromagnet by the injector metal via spin Hall effect, and magnon conductivity is quantified by electrically measuring the magnon density underneath the detector metal. The central metal bath is thermalized at temperature $T_F$ and may possess a spin-Hall generated nonequilibrium spin accumulation $\mu_s$ that can modulate the magnon density below it. The entire ferromagnet is coupled to a phonon bath at temperature $T_B$. The three ferromagnetic regions in between the injector and detector metals can be modeled as three resistances in series, as shown below the device. At the bottom of the figure, we sketch the external magnetic field, with a negative uniform gradient along the $x$ axis, considered in the spin conductivity calculation.}
\label{fig6}
\end{figure}
Spin conductivity predictions in Sec.~\ref{sigma} can be experimentally verified using a two-terminal spin transport setup similar to the devices studied in, e.g., Refs.~\onlinecite{cornelissenPRL18,wimmerCM18}. The setup is shown in Fig.~\ref{fig6}. The magnons are injected from the injector metal via the spin Hall effect, and magnon spin conductivity is quantified by electrically measuring the magnon density underneath the detector metal. The central metal bath is thermalized at temperature $T_F$ and may possess a nonequilibrium spin accumulation $\mu_s$ generated via the spin Hall effect. The main function of the spin accumulation is to modulate the density of magnons below the metal and to alter the spin conductivity in that region. The entire ferromagnet is coupled to a phonon bath at temperature $T_B$.

We characterize spin transport in between the injector and detector metals using a series resistor model (see Fig.~\ref{fig6}). Given that the cross-sectional area of the ferromagnet is $\sA$ and that the conductivities in the three regions from left to right are $\s'$, $\s_0$ and $\s'$, respectively, the total spin resistance $R=1/G$ of the region should be given by
\beq
\label{sres}
R=\frac{1}{G}=\frac{2d}{\s'\sA}+\frac{w}{\s_0\sA}\equiv2R'+R_0\ .
\eeq
The spin conductivity $\s'$ for the region outside of the central region (beneath the metal bath) can be obtained from $\s_0$ in Eq.~\eqref{s0} by setting $\al_F=0$, i.e.,
\beq
\frac{1}{R'}=\left.\frac{\s_0\sA}{d}\right|_{\al_F=0}\ .
\eeq

\begin{figure}[t]
\includegraphics[width=.9\linewidth]{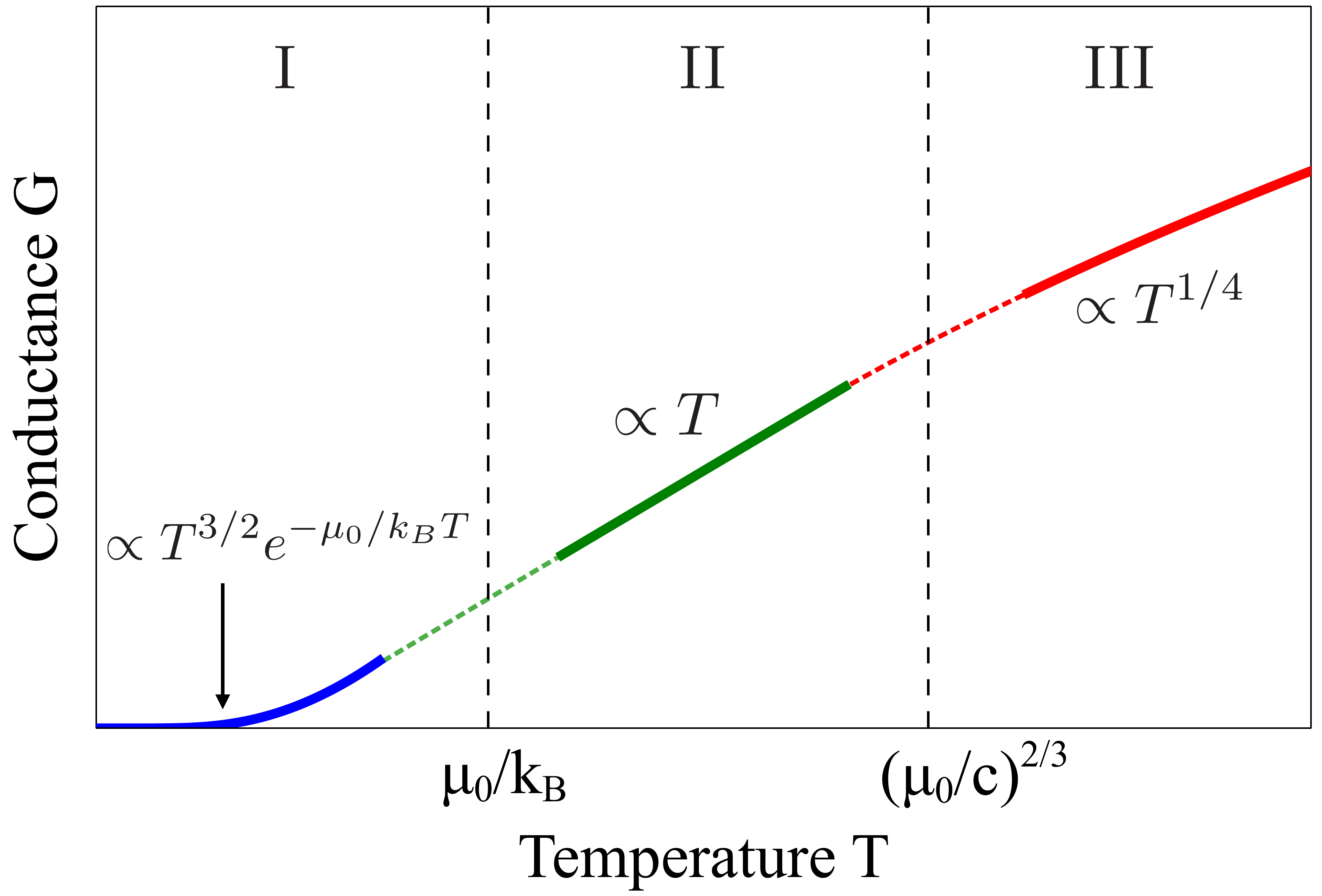}
\caption{(color online) The total spin conductance for the magnon system located in between the injector and detector metals for a fixed magnetic field $\mu_0>0$ and for $\mu_s=0$. We assume here that $T_B=T_F\equiv T$. The conductance traverses through three regimes as the bath temperature $T$ increased from the low temperature regime $T\ll\mu_0/k_B$ (regime I), through the intermediate regime $\mu_0/k_B\ll T\ll (\mu_0/c)^{3/2}$ (regime II), and finally to the high temperature regime $T\gg(\mu_0/c)^{3/2}$ (regime III).}
\label{fig7}
\end{figure}
The first set of experiments can be performed in the absence of electrical pumping, i.e., $\mu_s=0$, maintaining the temperatures of the two baths equal, i.e., $T_F=T_B=T$, and sweeping $T$. Here, the external field is fixed at some positive value with $\mu_0=\hbar\g B-SK_\bzero>0$, and the gradual increasing of $T$ from small ($T\ll\mu_0/k_B$) to large [$T\gg(\mu_0/c)^{3/2}$] values allows one to probe the finite temperature crossover behavior of the spin conductivity. The expected crossover behavior is shown in Fig.~\ref{fig7}. At small temperatures $T\ll\mu_0/k_B$ (regime I), the spin conductivities $\s',\s_0$ both scale as $T^{3/2}e^{-\mu_0/k_BT}$ [see Eq.~\eqref{sclow}] as shown by the blue line in Fig.~\ref{fig7}. As the temperature increases into the intermediate and high temperature regimes (see regimes II and III in Fig.~\ref{fig7}), the spin conductivities (and hence the spin conductance) exhibit power law behavior. In the intermediate temperature regime $\mu_0/k_B\ll T\ll(\mu_0/c)^{3/2}$, $\s'$ and $\s_0$ both scale linearly in $T$, while the behavior crosses over to $T^{1/4}$ in the high temperature regime $T\gg(\mu_0/c)^{3/2}$. 

\begin{figure}[t]
\includegraphics[width=\linewidth]{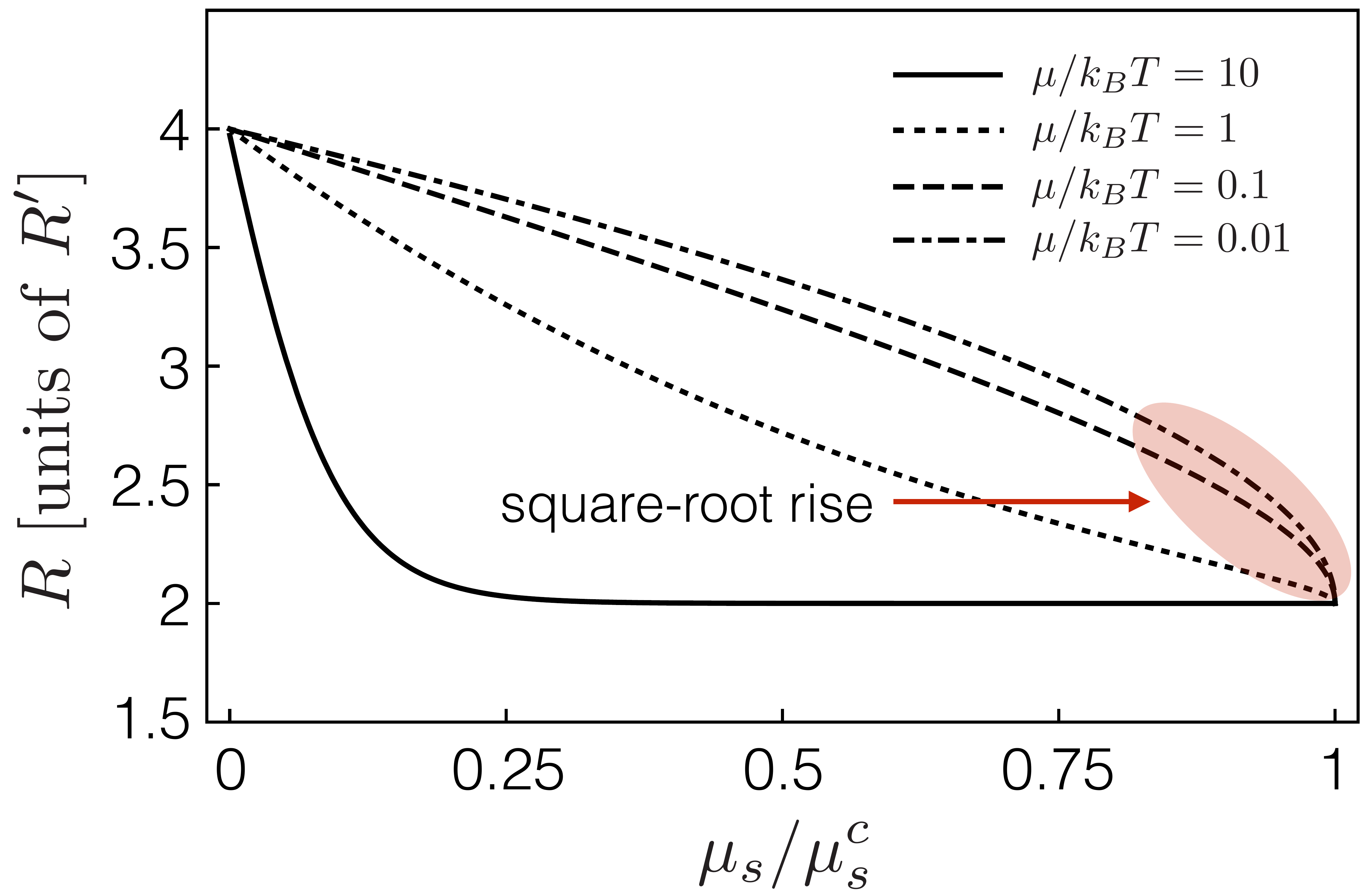}
\caption{(color online) Spin resistance $R$ as a function of electrical pumping $\mu_s$ for various values of $\mu/k_BT$, where $T_B=T_F\equiv T$. Here, $R$ is plotted in units of $R'$ and we have used $\al_B=\al_F=0.01$ and $d=w$.}
\label{fig8}
\end{figure}

We now include the effect of the electrical pumping $\mu_s$. Here, we fix the external magnetic field so that $\mu_0>0$, fix both bath temperatures to $T$, but sweep the nonequilibrium spin accumulation $\mu_s$ toward $\mu_s^c$. The spin resistance outside of the region modulated by the central metal bath is then fixed to a value $R'$ that is independent of $\mu_s$, but $1/R_0=\s_0\sA/w$ exhibits a diverging behavior as already shown in Sec.~\ref{scdrive}. In Fig.~\ref{fig8}, we plot the total spin resistance $R$ (in units of $R'$) as a function of $\mu_s$ for $d=w$ and for various ratios of $\mu/k_BT=0.01,0.1,1,10$. We find that as $\mu_s$ approaches the critical value, the resistance in the central region below the metal bath vanishes, so that the total spin resistance approaches $2R'$. The plots evince a square root rise in the region $\mu_s\lesssim\mu_s^c$ for relatively high temperatures, e.g., $\mu/k_BT\sim0.01$, while this behavior changes for relatively low temperatures (see, e.g., $\mu/k_BT=10$).

In Fig.~\ref{fig8}, we have chosen $\al_B=\al_F=0.01$. The $y$-intercepts of all the curves generally move up (down) when $\al_F$ is increased above (decreased below) $\al_B$; while the intercepts shift, the qualitative shapes of the curves do not change even when $\al_F$ deviates from $\al_B$.

\section{Conclusion}
\label{conc}
We present a microscopic theory based on the Keldysh path integral formalism to study a magnon system in contact with a phonon bath and a metallic bath, and subjected to dc electrical spin injection. For mismatched bath temperatures and/or finite dc electrical pumping, the magnon system converges to a nonthermal steady state in which the magnon distribution function is given by a nontrivial superposition of the bath distribution functions. Focusing exclusively on the normal (uncondensed) phase, we uncover a rich finite temperature crossover behavior exhibited by the correlation length associated with the superfluid order parameter fluctuations as a function of the phonon and metallic bath temperatures. Motivated by recent spin transport measurements on a magnon system close to a BEC instability,\cite{wimmerCM18} we compute the linear spin conductivity for arbitrary bath temperatures and electrical pumping strengths, and show that the finite temperature crossovers in the correlation length leads to nontrivial dependences of the spin conductivity on the bath temperatures. In the presence of pumping, we find an inverse square-root divergence in the spin conductivity as the pumping strength approaches the threshold value for BEC instability. A two-terminal spin transport setup capable of verifying our predictions is presented. 

Recently, the stability criteria for a magnon gas in contact with a metallic bath and in the presence of electrical spin current injection were studied using the Keldysh path integral formalism.\cite{troncosoPRB19} While the microscopic Keldysh approach was applied to the normal phase in both Ref.~\onlinecite{troncosoPRB19} and here, extending this approach for spin transport analysis in the condensed phase is an interesting future direction. A microscopic Keldysh formulation of the coupled nonlinear dynamics involving the condensate and the thermal magnons is also an interesting open problem.

Lastly, using the Keldysh formalism to understand the BEC of parametrically-pumped magnons would also be a worthwhile endeavor. However, a straightforward generalization is hampered, at least, by the fact that the parametrically-pumped magnon BEC is a transient (dynamic) phenomenon so that the current formalism must be extended to capture the nonequilibrium transient physics. Furthermore, understanding BEC formation following particle and energy injection into a narrow region of the spectrum requires one to understand how these excess particles and energy transverse through the energy shells after the pump is turned off. A well-suited formalism in treating this physics may be the theory of weak turbulence, which has been applied to various interacting Bose gases as well as semiconductor lasers.\cite{lvovPHYD98,nazarenkoPHYD06,promentPHYD12,zakharovBOOK84} It would be interesting to apply the idea of weak turbulence to magnon BEC in parametrically-pumped magnetic insulators.

\acknowledgments
The author would like to thank Yaroslav Tserkovnyak for a helpful correspondence. This research was supported by Research Foundation CUNY Project \#~90922-07 10.

\appendix
\section{Tracing out the boson bath}
\label{traceapp}
Performing the Gaussian integrals over the boson bath degrees of freedom in Eq.~\eqref{z} with Eq.~\eqref{sbi} directly leads to Eq.~\eqref{sbi2}, where $\S^{R,A,K}_\bi(\W)=\sum_\al|\ka_{\bi,\al}|^2d^{R,A,K}_\al(\W)$ are the phonon-induced magnon self-energies. The retarded component may be written as 
\beq
\label{sigr2}
\S^R_\bi(\W)=\int\frac{d\W'}{2\p}\frac{2J^B_\bi(\W')}{\W-\W'+\ii\de}\ ,
\eeq
where $J^B_\bi(\W)=\p\sum_\al|\ka_{\bi,\al}|^2\de(\W-\W_\al)$ is the spectral density of the $\bi$-th bath, and the Keldysh component reads
\beq
\S^K_\bi(\w)=-2\ii J^B_\bi(\w)\coth\round{\frac{\hbar\w}{2k_BT_B}}\ .
\eeq
Equation~\eqref{sigr2} shows that the current formalism allows one to consider any functional form for the phonon spectral density $J^B_\bi(\W)$. In this work, we consider an ohmic bath, 
\beq
J^B_\bi(\W)=\al^B_\bi\W\frac{\W_c^2}{\W_c^2+\W^2}\ ,
\eeq
where we have introduced a Drude cutoff function with cutoff frequency $\W_c$ and a site-dependent Gilbert damping parameter $\al^B_\bi$. In the low-frequency limit $\W\ll\W_c$, Eq.~\eqref{sigr2} reduces to 
\beq
\label{sigrexp}
\S^R_\bi(\W)=-\al^B_\bi(\W_c+\ii\W)\ .
\eeq
We therefore see that an ohmic bosonic bath here leads to the standard level broadening proportional to magnon frequency and thus to the familiar Gilbert damping phenomenology. In the main text, we have dropped the unimportant constant real part in Eq.~\eqref{sigrexp}.

\section{Tracing out the fermion bath}
\label{tracefapp}
Performing the Gaussian integrals over the fermion bath degrees of freedom in Eq.~\eqref{z} with Eq.~\eqref{smi} directly leads to 
\begin{multline}
\ii\cS^m_\bi=\tr\ln\Bigg[1-\frac{1}{\sqrt{2}}\sum_\bk\mat{\hg_{\bk\up}(\w)}{0}{0}{\hg_{\bk\down}(\w)}\\
\times\mat{0}{\eta_\bi\hat\cA_\bi(\w-\w')}{\eta^*_\bi\hat{\bar\cA}_\bi(\w-\w')}{0}\Bigg]\ ,
\end{multline}
where 
\beq
\hg_{\bk\s}(\w)=\mat{g^R_{\bk\s}(\w)}{g^K_{\bk\s}(\w)}{0}{g^A_{\bk\s}(\w)}\ ,\ \ \hat\cA_{\bi}(\w)=\mat{a^c_{\bi}(\w)}{a^q_{\bi}(\w)}{a^q_\bi(\w)}{a^c_{\bi}(\w)}\ .
\eeq
Expanding the $\tr\ln$ to second order in $\eta_\bi$ gives the Gaussian correction presented in Eq.~\eqref{smi2} of the main text, where the Keldysh components of the fermion-induced magnon self-energy matrix are given by
\begin{align}
\Pi^R_\bi(\W)&=-\frac{\ii|\eta_\bi|^2}{2}\sum_{\bk\bk'}\int\frac{d\w'}{2\p}\Big[g^R_\bk(\W+\w')g^K_{\bk'\down}(\w')\\
&\qquad\qquad\qquad+g^K_{\bk\up}(\W+\w')g^A_{\bk'}(\w')\Big]=\Pi^{A*}_\bi(\W)\\
\Pi^K_\bi(\W)&=-\frac{\ii|\eta_\bi|^2}{2}\sum_{\bk\bk'}\int\frac{d\w'}{2\p}\Big[g^K_{\bk\up}(\W+\w')g^K_{\bk'\down}(\w')\\
&\quad+g^R_\bk(\W+\w')g^A_{\bk'}(\w')+g^A_\bk(\W+\w')g^R_{\bk'}(\w')\Big]\ .
\end{align}
Inserting Eqs.~\eqref{fbgf} and performing the internal frequency integral immediately gives Eqs.~\eqref{pir} and \eqref{pik}.

The fourth order term in the $\tr\ln$ expansion gives rise to the following correction to the quartic terms in $\cS_F$ [i.e., the last two terms in Eq.~\eqref{sf}]
\begin{multline}
\ii\cS^{m(4)}_\bi=-\frac{|\eta_\bi|^4}{8}\sum_{\{\bk_i\}}\int\frac{d\w}{2\p}\int\frac{d\w_1}{2\p}\int\frac{d\w_2}{2\p}\int\frac{d\w_3}{2\p}\\
\times\tr\Big[\hg_{\bk_1\up}(\w+\w_1+\w_3-\w_2)\hat\cA_\bi(\w)\hg_{\bk_2\down}(\w_1+\w_3-\w_2)\\
\times\hat\cA^*_\bi(\w_2)\hg_{\bk_3\up}(\w_1+\w_3)\hat\cA_\bi(\w)\hg_{\bk_4\down}(\w_3)\hat\cA^*_\bi(\w+\w_1-\w_2)\Big]\ .
\end{multline}
Since we expect the energies of the magnons to be much smaller than the Fermi energy, we approximate the above expression by setting $\w=\w_1=\w_2=0$ in the arguments for the fermionic Green functions, and obtain
\begin{multline}
\ii\cS^{m(4)}_\bi=-\frac{|\eta_\bi|^4}{8}\int dt\sum_{\{\bk_i\}}\int\frac{d\w}{2\p}\\
\times\tr\Big[\hg_{\bk_1\up}(\w)\hat\cA_\bi(t)\hg_{\bk_2\down}(\w)\hat\cA^*_\bi(t)\hg_{\bk_3\up}(\w)\hat\cA_\bi(t)\hg_{\bk_4\down}(\w)\hat\cA^*_\bi(t)\Big]\ .
\end{multline}
For the retarded and advanced components, we have $\sum_\bk(\w-\ve_\bk/\hbar\pm\ii\de)^{-1}\approx-\ii\p\hbar\rho_0$, where we assume that the real parts give zero. Then performing the $\w$-integral, the fermions generate the following eight quartic terms in the effective magnon action
\begin{multline}
\label{dissq}
\ii\cS^{m(4)}_\bi=-\ii\round{\al^F_\bi}^2\int dt\Big[u_{1}a_\bi^{*c}a_\bi^{*c}a_\bi^{c}a_\bi^q-u_{1}a_\bi^{*c}a_\bi^{*q}a_\bi^{c}a_\bi^{c}\\
+u_{2}a_\bi^{*q}a_\bi^{*q}a_\bi^{c}a_\bi^{q}-u_{2}a_\bi^{*c}a_\bi^{*q}a_\bi^{q}a_\bi^{q}+u_{3}a_\bi^{*c}a_\bi^{*c}a_\bi^{q}a_\bi^q\\
+u_{3}a_\bi^{*q}a_\bi^{*q}a_\bi^{c}a_\bi^{c}+u_{4}a_\bi^{*c}a_\bi^{*q}a_\bi^{c}a_\bi^{q}+u_{5}a_\bi^{*q}a_\bi^{*q}a_\bi^{q}a_\bi^{q}\Big]\ ,
\end{multline}
where
\begin{align}
u_{1}&=\p \ii\frac{k_BT_F}{\hbar}\frac{\mu_s}{k_BT_F}\\
u_{2}&=\p \ii\frac{k_BT_F}{\hbar}\frac{5\frac{\mu_s}{k_BT_F}+3\frac{\mu_s}{k_BT_F}\cosh\round{\frac{\mu_s}{k_BT_F}}-8\sinh\round{\frac{\mu_s}{k_BT_F}}}{\cosh\round{\frac{\mu_s}{k_BT_F}}-1}\\
u_{3}&=\p \ii\frac{k_BT_F}{\hbar}\square{2-\frac{\mu_s}{k_BT_F}\coth\round{\frac{\mu_s}{2k_BT_F}}}\\
u_{4}&=-4\p \ii\frac{k_BT_F}{\hbar}\square{1-\frac{\mu_s}{k_BT_F}\coth\round{\frac{\mu_s}{2k_BT_F}}}\\
u_{5}&=-\p \ii\frac{k_BT_F}{\hbar}\coth\round{\frac{\mu_s}{2k_BT_F}}\csch^2\round{\frac{\mu_s}{2k_BT_F}}\\
&\qquad\times\square{3\frac{\mu_s}{k_BT_F}+\frac{\mu_s}{k_BT_F}\cosh\round{\frac{\mu_s}{k_BT_F}}-4\sinh\round{\frac{\mu_s}{k_BT_F}}}\ .
\end{align}
While the first 4 terms in Eq.~\eqref{dissq} renormalize the existing vertices in Eq.~\eqref{sf}, the remaining terms are new quartic vertices generated by the integration over the bath.

\section{Self-consistent equation for the magnon gap for $\mu_s\ne0$}
\label{neqscmu}
In this appendix, we consider Eq.~\eqref{mass} in the presence of the nonequilibrium drive $\mu_s$, 
\begin{multline}
\label{sc2}
\mu=\mu_0+\Bigg[\mathfrak{M}_B\round{\frac{\mu}{k_BT_B},\frac{\mu_s}{k_BT_B}}\\
+\mathfrak{M}_F\round{\frac{\mu}{k_BT_F},\frac{\mu_s}{k_BT_F}}\Bigg]cT^{3/2}\ ,
\end{multline}
where the two dimensionless functions are given by
\begin{align}
\mathfrak{M}_B(x,y)&=\frac{2\al_B}{\al\sqrt{\p}\z_{3/2}(1)}\int_0^\infty\frac{ds\sqrt{s}}{s+x-(\al_F/\al)y}\frac{s+x}{e^{s+x}-1}\ ,\\
\mathfrak{M}_F(x,y)&=\frac{2\al_F}{\al\sqrt{\p}\z_{3/2}(1)}\int_0^\infty\frac{ds\sqrt{s}}{s+x-(\al_F/\al)y}\frac{s+x-y}{e^{s+x-y}-1}\ .
\end{align}
If we numerically solve Eq.~\eqref{sc2} for $\mu$ for a given set of $T_B$ and $T_F$, we find that a finite $\mu_s>0$ gives relatively small corrections to the magnon gap obtained for the same set of bath temperatures and $\mu_s=0$. This allows us to approximate $\mu$, even with finite electrical pumping, by setting $\mu_s=0$ in Eq.~\eqref{mass} and treat $\mu_s$ as an independent nonequilibrium parameter.

\section{Kubo formula for the spin conductivity}
\label{kubo}
In this appendix, we derive the expression for the dc spin conductivity given in Eq.~\eqref{s0} of the main text. We start with the Fourier transformed expression for the spin conductivity $\s(\bq,\w)=\chi(\bq,\w)/(-\ii q_x)$, where
\beq
\label{chi1}
\chi(\bq,\w)=-\frac{\ii}{\hbar\sV}\int_0^\infty dt\ e^{\ii\w t}\langle [j(\bq,t),\varrho(-\bq,0)]\rangle_{\rm HF}\ .
\eeq
The spin current and magnon density operators $j$ and $\varrho$ have been introduced in the main text. Integrating Eq.~\eqref{chi1} by parts and using the relation $\hbar\dot\varrho(\bq,t)=-\ii\bq\cdot\bj(\bq,t)$, we obtain
\begin{multline}
\chi(\bq,\w)=-\frac{\ii}{\hbar\sV}\frac{1}{\ii\w}\Bigg\{-\langle [j(\bq,t),\varrho(-\bq,0)]\rangle_{\rm HF}\\
+\frac{\ii q_x}{\hbar}\int_0^\infty dt\ e^{\ii\w t}\langle [j(\bq,t),j(-\bq,0)]\rangle_{\rm HF}\Bigg\}\ .
\end{multline}
Evaluating the above expectation values within the Hartree-Fock approximation, we then obtain
\beq
\chi(\bq,\w)=-\frac{\ii q_x}{\w\sV}\square{\frac{2\Xi}{\hbar}\sum_\bk\int\frac{d\W}{2\p}\sD^<_\bk(\W)+\round{\frac{\Xi}{\hbar}}^2P(\w)}\ ,
\eeq
where $\Xi\equiv J_\bzero S\xi^2_J/2$, $\sD^<_\bk(\W)=-\ii N(\W)\cB_\bk(\W)$ is the lesser magnon Green function, and
\begin{multline}
\label{pomega}
P(\w)=\int\frac{d\W}{2\p}\sum_\bk(2k_x)^2\Big[\sD^R_{\bk}(\W+\w)\sD^<_{\bk}(\W)\\
+\sD^<_{\bk}(\W+\w)\sD^A_{\bk}(\W)\Big]\ .
\end{multline}

We find that the $\w=0$ contribution of the second term in $\chi(\bq,\w)$ cancels precisely the first term. We may therefore write the spin conductivity in the uniform limit as
\beq
\s(\bq\rightarrow0,\w)=-\frac{1}{\sV}\round{\frac{J_\bzero S\xi_J^2}{2\hbar}}^2\frac{P(\w)-P(0)}{\w}\ .
\eeq
Using Eq.~\eqref{pomega} and taking the limit $\w\rightarrow0$ in the above expression immediately gives Eq.~\eqref{s0} in the main text.

%

\end{document}